\documentclass[article,preprint,12pt,unsrt,compress]{elsarticle}
\usepackage{a4wide}
\usepackage{graphicx}
\usepackage{amsfonts}
\usepackage{amsmath}
\usepackage{booktabs}
\usepackage{epsfig}
\usepackage{stmaryrd}
\usepackage{hyperref}
\hypersetup{citecolor=black,linkcolor= black}

\begin{document}
\title{Power-law cross-correlations estimation under heavy tails}
\author{Ladislav Kristoufek}
\ead{kristouf@utia.cas.cz}
\address{Institute of Information Theory and Automation, The Czech Academy of Sciences, Pod Vodarenskou Vezi 4, 182 08, Prague 8, Czech Republic\\
Institute of Economic Studies, Faculty of Social Sciences, Charles University, Opletalova 26, 110 00, Prague 1, Czech Republic
}

\begin{abstract}
We examine the performance of six estimators of the power-law cross-correlations -- the detrended cross-correlation analysis, the detrending moving-average cross-correlation analysis, the height cross-correlation analysis, the averaged periodogram estimator, the cross-periodogram estimator and the local cross-Whittle estimator -- under heavy-tailed distributions. The selection of estimators allows to separate these into the time and frequency domain estimators. By varying the characteristic exponent of the $\alpha$-stable distributions which controls the tails behavior, we report several interesting findings. First, the frequency domain estimators are practically unaffected by heavy tails bias-wise. Second, the time domain estimators are upward biased for heavy tails but they have lower estimator variance than the other group for short series. Third, specific estimators are more appropriate depending on distributional properties and length of the analyzed series. In addition, we provide a discussion of implications of these results for empirical applications as well as theoretical explanations.
\end{abstract}


\journal{Communications in Nonlinear Science and Numerical Simulation}

\maketitle
%
%

\newpage

\section{Introduction}


Power-law cross-correlations have brought a new perspective into analysis of bivariate time series with applications across a wide range of disciplines -- hydrology \cite{Hajian2010}, (hydro)meteorology \cite{Vassoler2012,Kang2013}, seismology and geophysics \cite{Shadkhoo2009,Marinho2013}, economics and finance \cite{Podobnik2009a,He2011,He2011a,Lin2012,Cao2012,Shi2014,Zhao2014}, biometrics \cite{Ursilean2009}, biology \cite{Xue2012}, DNA sequences \cite{Stan2013}, neuroscience \cite{Jun2012}, electricity \cite{Wang2013}, traffic \cite{Zebende2009,Xu2010,Zhao2011,Yin2015}, and others. Analyzing cross-correlations between a pair of time series brings new insights into their dynamics and specifically, the power-law behavior of these may suggest very special features compared to exponential (vanishing) cross-correlations \cite{Kantelhardt2009,Horvatic2011,Podobnik2011,Sela2012,Kristoufek2015,Kristoufek2015b}. Presence of power-law cross-correlations is a very current topic in various scientific fields but, unfortunately, empirical papers strongly outnumber theoretical ones discussing their emergence and origin. Podobnik \textit{et al.} \cite{Podobnik2008a} suggest that power-law cross-correlated processes can occur as a mixture of correlated long-range correlated processes. Sela \& Hurvich \cite{Sela2012} and Kristoufek \cite{Kristoufek2015,Kristoufek2015b} add a possibility of long-range correlated processes with cross-correlated error terms as sources of long-range cross-correlations together with a discussion of various properties of such processes. Investigation of power-law cross-correlated processes has also led to an important novelty in analyzing cross-correlations at specific scales \cite{Zebende2011,Yin2013,Kristoufek2014a,Kristoufek2014b} as well as a new regression framework for specific scales \cite{Kristoufek2015c}.

Formally, processes labeled as power-law (long-range, long-term) cross-correlated can be defined in both time and frequency domain. In the time domain, the long-range cross-correlated processes $\{x_t\}$ and $\{y_t\}$ are characterized by a power-law (hyperbolically) decaying cross-correlation function $\rho_{xy}(k)$ with a time lag $k$ so that $\rho_{xy}(k)\propto k^{2H_{xy}-2}$ for $k \rightarrow +\infty$ \cite{Kristoufek2013}. In the frequency domain, the processes have a divergent at origin cross-power spectrum. Specifically, the cross-power spectrum $f_{xy}(\lambda)$ with frequency $\lambda$ scales as $|f_{xy}(\lambda)|\propto \lambda^{1-2H_{xy}}$ for $\lambda \rightarrow 0+$ \cite{Sela2012}. The bivariate Hurst exponent $H_{xy}$ is a measure of long-range cross-correlations, specifically their decay. For $H_{xy}=0.5$, the processes are not considered long-range cross-correlated whereas for $H_{xy}>0.5$, these are referred to as the cross-persistent processes which tend to move together. Time series with $H_{xy}<0.5$ form a very specific type of processes which have been theoretically only sparsely examined.

In the same way as for the univariate series, the correlations scaling and the underlying distribution tail behavior are tightly interconnected \cite{Taqqu1995,Barunik2010}. Investigation of the relationship between the two goes back to Mandelbrot \& Wallis \cite{Mandelbrot1968} -- and thus hydrology -- when examining the emerge of the so-called Hurst effect \cite{Hurst1951}. Even though the effect emerges both for long-range dependent processes and for processes with heavy tails, it is the correlation function, and specifically its shape, which is crucial. The confusion about the real and the spurious sources of the Hurst effect spreads also into self-similarity and phase transitions \cite{Samorodnitsky2006}. Such spurious effects can easily transfer into bivariate, or in general multivariate, setting which forms a motivational cornerstone of this text.

Presence of heavy tails in distributions, or in other words high occurrence of extreme events, is well documented across disciplines \cite{Adler1998,Cont2001,Egozcue2001,Katz2002,Bondar2004,Hernandez-Campos2004,Barabasi2005,Vazquez2006,ElAdlouni2008,Pisarenko2010,Reynolds2011}. However, the current stream of literature does not take a possible effect on estimators of the bivariate Hurst exponent implied by such heavy-tailed (fat-tailed) distributions into consideration. Here we fill this gap by focusing on three frequency domain estimators -- the average periodogram estimator \cite{Sela2012}, the cross-periodogram estimator and the local cross-Whittle estimator \cite{Kristoufek2014} -- and three time domain estimators -- the detrended cross-correlation analysis \cite{Podobnik2008}, the detrending moving-average cross-correlation analysis \cite{He2011,Arianos2009} and the height cross-correlation analysis \cite{Kristoufek2011} -- performance of which we examine under such distributions. Specifically, we study the effect of varying characteristic exponent of the $\alpha$-stable distributions on bias, variance and mean squared error of the estimators for various time series lengths.

\section{Methods} 

In this section, we shortly describe six estimators of the bivariate Hurst exponent. These are separated into two groups based on their domain of operation -- time and frequency. The simulations setting is then described in detail.

\subsection{Time domain estimators}

The detrended cross-correlation analysis (DCCA, or DXA) \cite{Podobnik2008} is the most popular time domain method of the bivariate Hurst exponent estimation, constructed as a bivariate generalization of the detrended fluctuation analysis (DFA) \cite{Peng1993,Peng1994}. The method has led to various generalizations and expansions \cite{Zhou2008,Gu2010,Jiang2011}. Considering two long-range cross-correlated series $\{x_t\}$ and $\{y_t\}$ with $t=1,\ldots,T$ and their respective profiles $\{X_t\}$ and $\{Y_t\}$, the DCCA procedure is based on an examination of the detrending covariances scaling with respect to scale $s$. Specifically, the time series are divided into overlapping boxes of length $s$ and the linear time trend is estimated in each box yielding $\widehat{X_{k,j}}$ and $\widehat{Y_{k,j}}$ for boxes $j\le k \le j+s-1$. The detrended covariance $f_{DCCA}^2(s,j)$ is obtained for each box $j$ of length $s$ and it is further averaged over all boxes of length $s$ to get $F^2_{DCCA}(s)$ as an estimated covariance for scale $s$. For the power-law cross-correlated processes, we have $F_{DCCA}^2(s)\propto s^{2H_{xy}}$. There are various ways of treating overlapping and non-overlapping boxes for scales $s$ as the method can become computationally highly demanding \cite{Taqqu1995,Barunik2010,Kristoufek2010,Grech2013,Grech2013a}. Due to this fact, we use non-overlapping boxes with a minimum scale of 10, a maximum scale of $T/5$ and a step between $s$ equal to 10 in the simulations.

The height cross-correlation analysis (HXA) \cite{Kristoufek2011} is a bivariate generalization of the height-height correlation analysis (HHCA) \cite{Barabasi1991a,Barabasi1991b} and the generalized Hurst exponent approach (GHE) \cite{DiMatteo2003,DiMatteo2005,DiMatteo2007}. The method is based on scaling of the height-height covariance function $K_{xy,2}(\tau)=\frac{\nu}{T^{\ast}}\sum_{t=1}^{T^{\ast}/\nu}|\Delta_{\tau}X_tY_t| \equiv \langle|\Delta_{\tau}X_tY_t|\rangle$ of detrended profiles $\{X_t\}$ and $\{Y_t\}$ with time resolution $\nu$ and $t=\nu,2\nu,...,\nu\lfloor\frac{T}{\nu}\rfloor$, where $\lfloor \rfloor$ is a lower integer sign. We denote $T^{\ast}=\nu\lfloor\frac{T}{\nu}\rfloor$, which varies with $\nu$, and we write the $\tau$-lag difference as $\Delta_{\tau}X_t \equiv X_{t+\tau}-X_t$ and $\Delta_{\tau}X_tY_t \equiv \Delta_{\tau}X_t\Delta_{\tau}Y_t$ for better legibility. The covariance function scales as $K_{xy,2}(\tau) \propto \tau^{H_{xy}}$ for the power-law cross-correlated processes. Di Matteo \textit{et al.} \cite{DiMatteo2003,DiMatteo2005,DiMatteo2007} suggest using the jackknife method to obtain more precise estimates of the Hurst exponent. In our simulations, we set $\tau_{min}=1$ and vary $\tau_{max}=5,\ldots,20$ to get the estimated $H_{xy}$ as an average of these.

The detrended moving-average cross-correlation analysis (DMCA) \cite{He2011,Arianos2009} is a generalization of the detrending moving average (DMA) \cite{Vandewalle1998,Alessio2002}. The method is in a way similar to DCCA as it assumes the power-law scaling of detrended covariances. However, there is no box-splitting involved in the procedure, which makes DMCA much less computationally demanding. Specifically, the profiles $\{X_t\}$ and $\{Y_t\}$ are detrended using the moving average of length $\kappa$ and the covariances of residuals $F_{DMCA}^2(\kappa)$ scale as $F_{DMCA}^2(\kappa)\propto \kappa^{2H_{xy}}$ for the power-law cross-correlated processes. In our setting, we use the non-weighted centered moving averages with $\kappa_{min}=11$ and $\kappa_{max}=1+T/5$ with a step of 2, i.e. parallel to the DCCA setting.

\subsection{Frequency domain estimators}

Construction of the frequency domain estimators is based on the divergence at origin of the cross power-spectrum of the power-law cross-correlated processes. As the cross-power spectrum is unobservable, its estimation becomes a crucial part of all the frequency domain procedures. For this purpose, a cross-periodogram $I_{xy}(\lambda)$ is the simplest and the most popular tool. It is defined as
\begin{multline}
\label{eq:cross-periodogram}
I_{xy}(\lambda_j)=\frac{1}{2\pi}\sum_{k=-\infty}^{+\infty}{\widehat{\gamma}_{xy}(k)\exp(-i\lambda_jk)}=\frac{1}{2\pi T}\sum_{t=1}^{T}{x_t\exp(-i\lambda_j t)}\sum_{t=1}^{T}{y_t\exp(i\lambda_j t)}=\\
I_x(\lambda_j)\overline{I_y(\lambda_j)},
\end{multline}
where $T$ is the time series length, $\widehat{\gamma}_{xy}(k)$ is an estimated cross-covariance at lag $k$, and $\lambda_j$ is a frequency defined as $\lambda_j=2\pi j/T$, where $j=1,2,\ldots,\lfloor T/2 \rfloor$ and $\lfloor \rfloor$ is the nearest lower integer operator. The cross-periodogram is thus defined between 0 and $\pi$ only. $I_x(\lambda_j)$ is a periodogram of series $\{x_t\}$, and $\overline{I_y(\lambda_j)}$ is a complex conjugate of a periodogram of series $\{y_t\}$.  To overcome the inconsistency of the raw periodogram in Eq. \ref{eq:cross-periodogram} \cite{Wei2006}, we use a smoothing operator based on Daniell \cite{Daniell1946,Bloomfield2000}. The smoothed cross-periodogram is utilized during the following estimation procedures.

Sela \& Hurvich \cite{Sela2012} propose the averaged periodogram estimator (APE) as a bivariate generalization of the method of Robinson \cite{Robinson1994}. Using a cumulative cross-periodogram 
\begin{equation}
\widehat{F_{xy}}(\lambda)=\frac{2\pi}{m}\sum_{j=1}^{\lfloor m\lambda/2\pi \rfloor}{I_{xy}(\lambda_j)},
\end{equation}
where $m\le T/2$ is a bandwidth parameter and fixed $q\in (0,1)$, the estimator is given as
\begin{equation}
\widehat{H_{xy}}=1-\frac{\log\frac{\widehat{F_{xy}}(q\lambda_m)}{\widehat{F_{xy}}(\lambda_m)}}{2\log q}.
\end{equation}
We follow the suggestion of Sela \& Hurvich \cite{Sela2012} and use $q=0.5$. Under numerous assumptions, the estimator is consistent, which is supported by simulations \cite{Sela2012,Kristoufek2014}.

The cross-periodogram estimator (XPE) proposed by Kristoufek \cite{Kristoufek2014} is based on the power-law scaling of cross power-spectrum close to the origin with a specific slope of $1-2H_{xy}$. The estimate of the bivariate Hurst exponent is obtained from
\begin{equation}
\label{eq:CP_H}
|I_{xy}(\lambda_j)| \propto \lambda_j^{1-2H_{xy}}.
\end{equation}
The scaling is expected to hold only for $\lambda \rightarrow 0+$ so that the regression is not performed over all frequencies. By choosing $\lambda_j=2\pi j/T$ for $j=1,2,\ldots,m$ where $m\le T/2$, the bivariate Hurst exponent is estimated using only the information up to the selected frequency specified by the bandwidth parameter $m$. The selection of $m$ influences the statistical properties of the estimator as an inclusion of more frequencies lowers variance but increases bias of the estimator. For the presented simulations, we follow the results of Kristoufek \cite{Kristoufek2014} and use $m=T/10$ for XPE as well as for the last analyzed estimator.

Kristoufek \cite{Kristoufek2014} introduces the local cross-Whittle estimator (LXW) as a bivariate generalization of the local Whittle estimator \cite{Robinson1995a} which is a semi-parametric maximum likelihood estimator based on the penalty function of K\"unsch \cite{Kunsch1987}. Specifically, the estimate of $H_{xy}$ is given as
\begin{equation}
\label{eq:LWX}
\widehat{H_{xy}}=\arg \min_{\frac{1}{2}<H_{xy}\le 1} R(H_{xy}),
\end{equation} 
where 
\begin{equation}
\label{eq:LWX_R}
R(H_{xy})=\log\left(\frac{1}{m}\sum_{j=1}^m{\lambda_j^{2H_{xy}-1}|I_{xy}(\lambda_j)|}\right)-\frac{2H_{xy}-1}{m}\sum_{j=1}^m{\log \lambda_j}
\end{equation}
and $\lambda_j=2\pi j/T$.

\subsection{Simulations setting}

We are interested in the effect of heavy tails on the power-law cross-correlations estimators performance. As a representative of heavy-tailed distributions, we select the $\alpha$-stable distributions which provide a rich parametric flexibility. We use the parametrization of Nolan \cite{Nolan2003}. The $\alpha$-stable distributions are characterized by four parameters -- $\alpha$, $\beta$, $\gamma$ and $\delta$. The heaviness of tails is represented by the characteristic exponent $0<\alpha\le 2$. Skewness is specified by parameter $-1\le \beta \le 1$ with $\beta=0$ signifying symmetry, $\beta>0$ positive skewness and $\beta<0$ negative skewness. Parameters $\gamma$ and $\delta$ describe scale and location of the distribution. The $\alpha$-stable distributed random variable with $S(\alpha,\beta,\gamma,\delta)$ has the following characteristic function:
  \begin{equation}
 \footnotesize
\label{eq2}
\phi(u)=
\left\{
\begin{array}{lr}
\exp(-\gamma^{\alpha}|u|^{\alpha}[1+i\beta(\tan\frac{\pi\alpha}{2})($sign$ u)(|\gamma u|^{1-\alpha}-1)]+i\delta u) & \alpha \ne 1 \\
\exp(-\gamma |u| [1+i\beta\frac{2}{\pi}($sign$ u) \ln (\gamma |u|)]+i\delta u) & \alpha = 1
\end{array}
\right.\end{equation}\\
There are only three cases with a closed-form density -- Gaussian ($\alpha=2$, $\beta=0$), Cauchy ($\alpha=1$, $\beta=0$), and L\'evy ($\alpha=1/2$, $\beta=1$) distributions. Importantly, the characteristic parameter $\alpha$ also specifies the existing moments of the distribution. For $\alpha$-stable distributions, only moments below $\alpha$ exist. Therefore, we have nonexistent variance for $\alpha<2$ and even nonexistent mean for $\alpha<1$.

In the simulations, we focus on the effect of the characteristic parameter $\alpha$, which manipulates the distribution tails,  on performance of six estimators of power-law cross-correlations. To do so, we follow two settings. For the first setting, one of the processes is the standard Gaussian noise and the other is the $\alpha$-stable random variable. For the latter process, we vary the parameter $\alpha$ between 1.1 and 2 with a step of 0.1. The other parameters are fixed to $\beta=0$, $\gamma=\sqrt{2}/2$ and $\delta=0$. The simulated series are standardized to have a unit variance. For the second setting, both processes are $\alpha$-stable distributed with the same $\alpha$ parameter.  For each parameter setting, we simulate 1000 series with time series lengths of $T=500,1000,5000,10000$ to see how the statistical properties of the estimators depend on the number of observations. The studied series are independent so that the expected bivariate Hurst exponent is equal to 0.5. As measures of performance, we study bias, standard deviation and mean squared errors of the estimators.

\section{Results and Discussion}

\begin{figure}[!htbp]
\begin{center}
\begin{tabular}{cc}
\includegraphics[width=75mm]{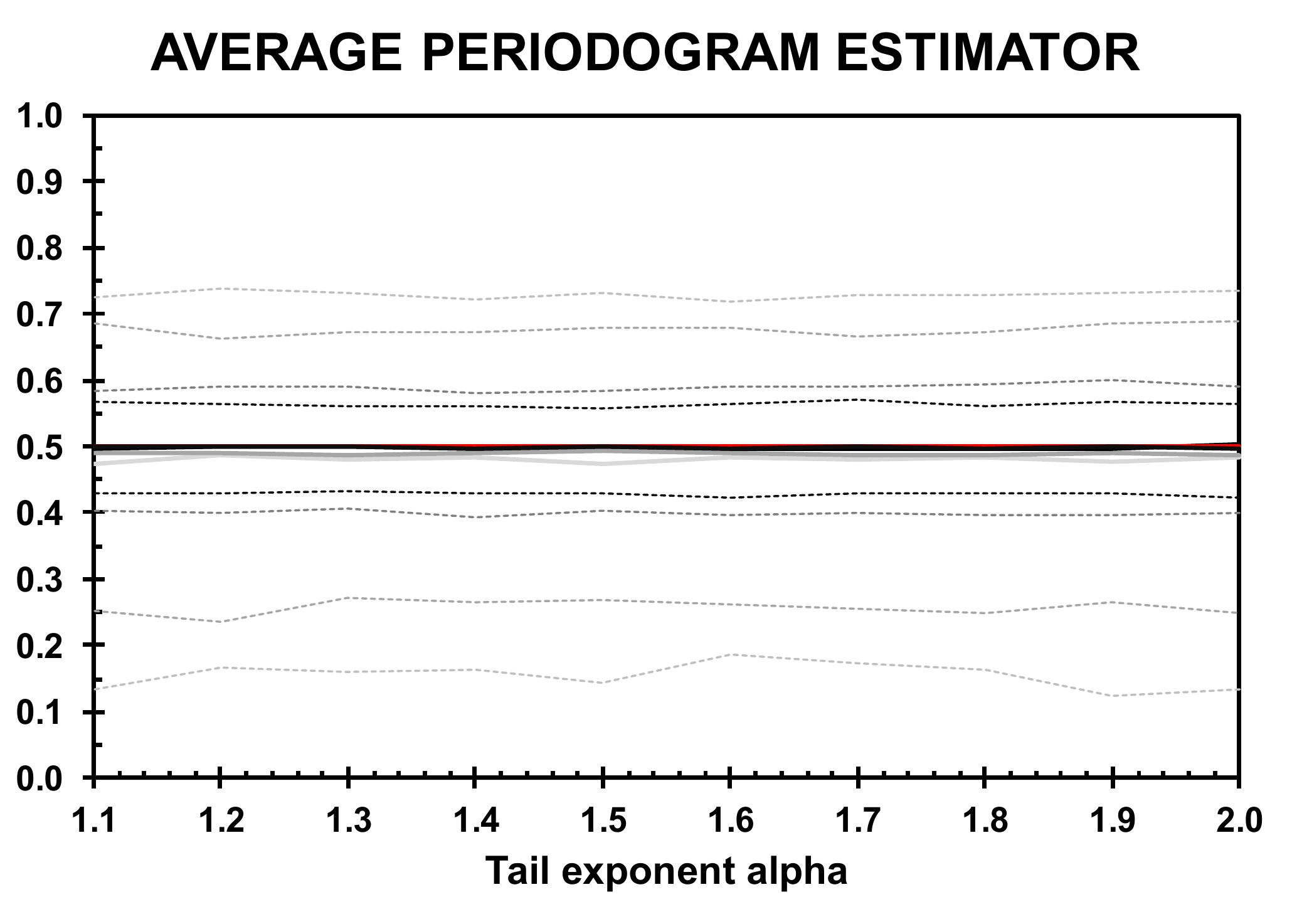}&\includegraphics[width=75mm]{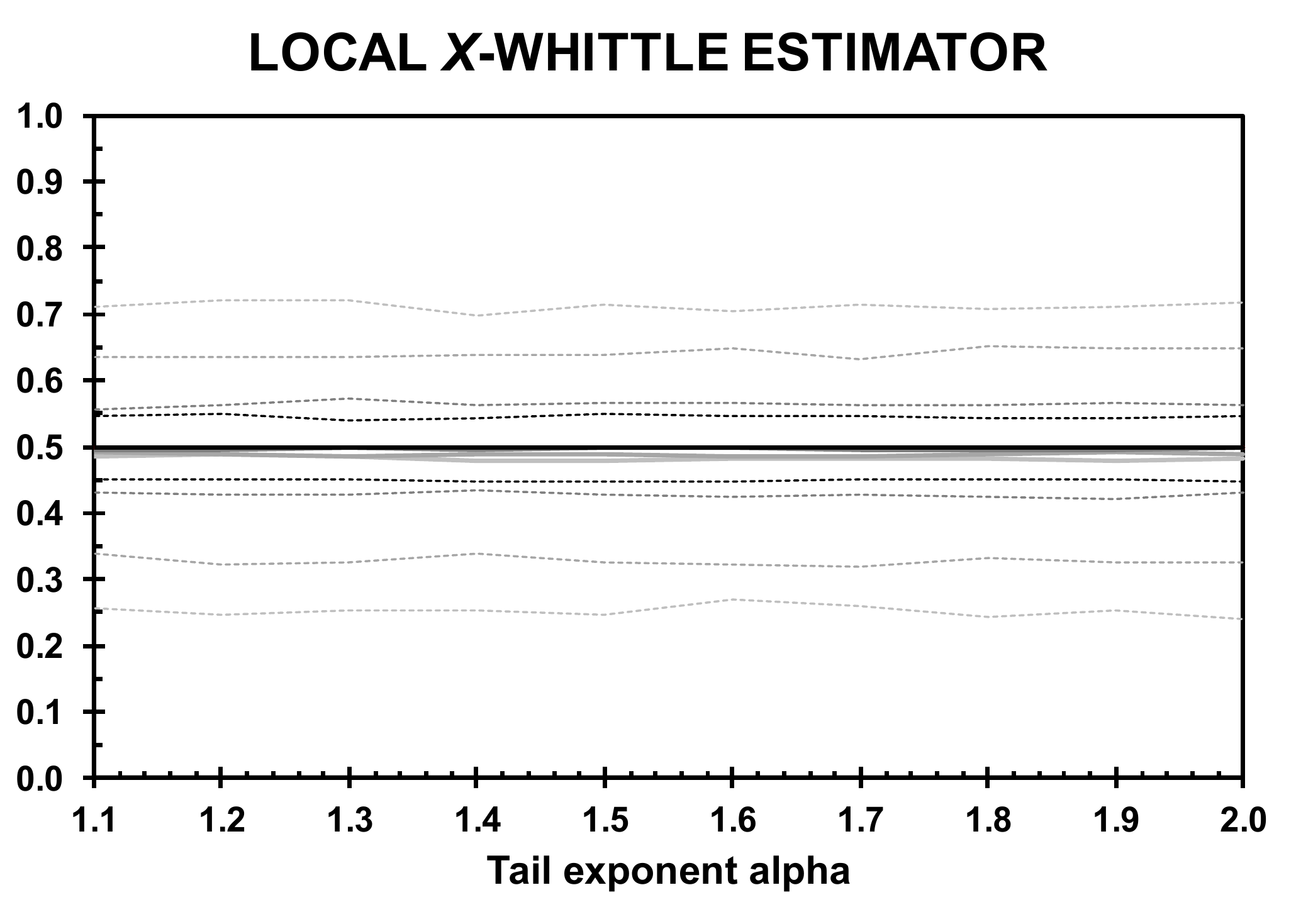}\\
\includegraphics[width=75mm]{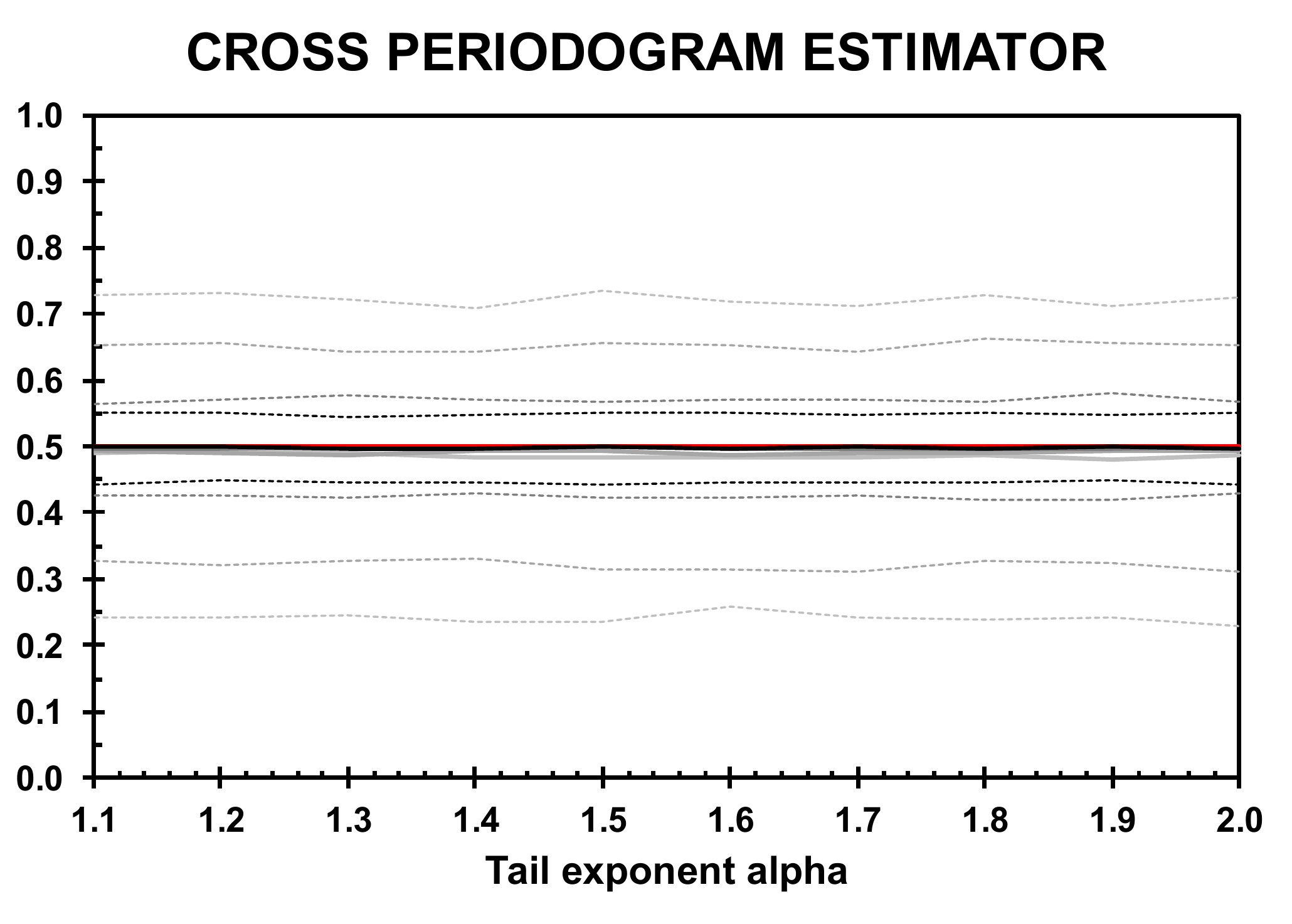}&\includegraphics[width=75mm]{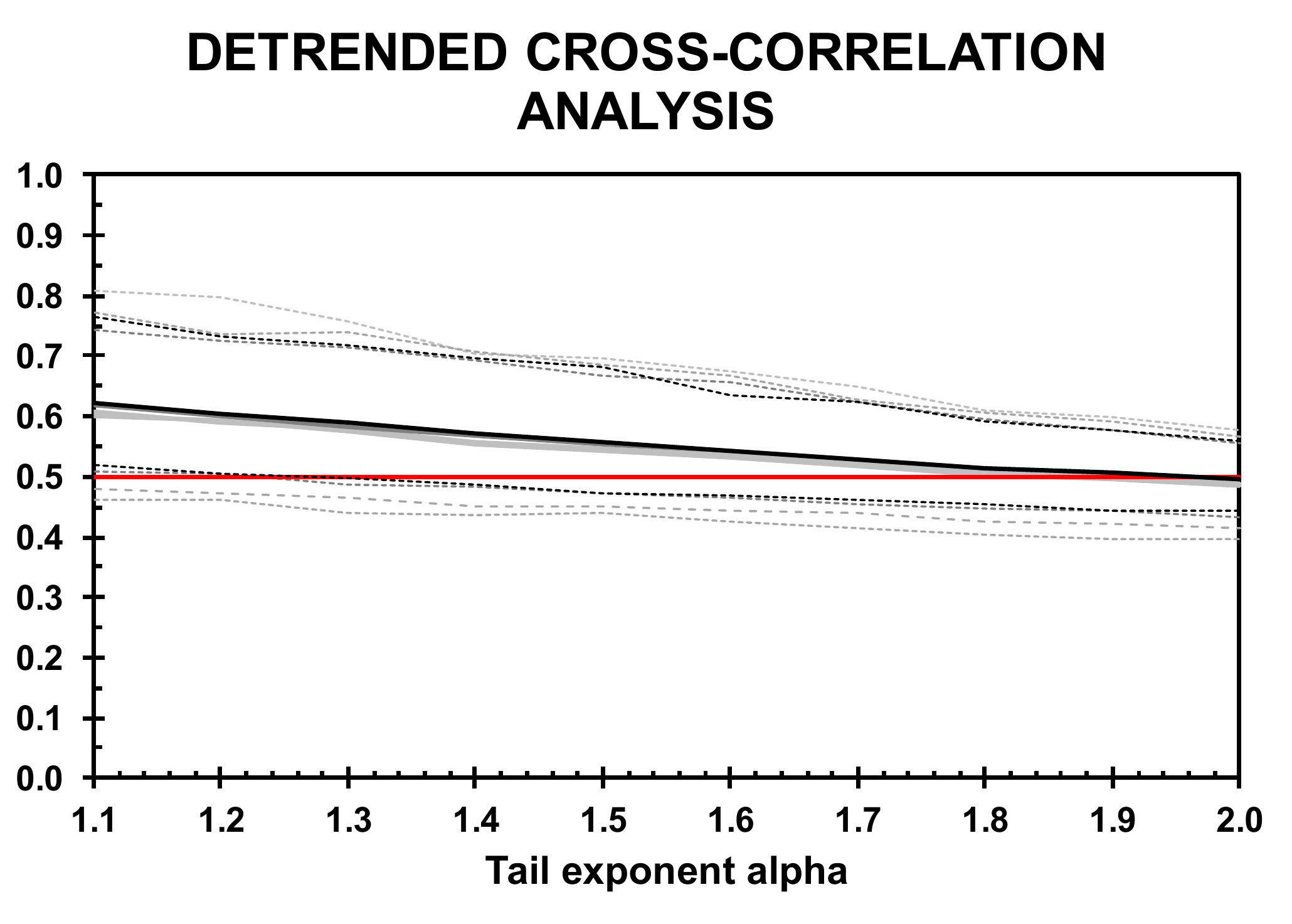}\\
\includegraphics[width=75mm]{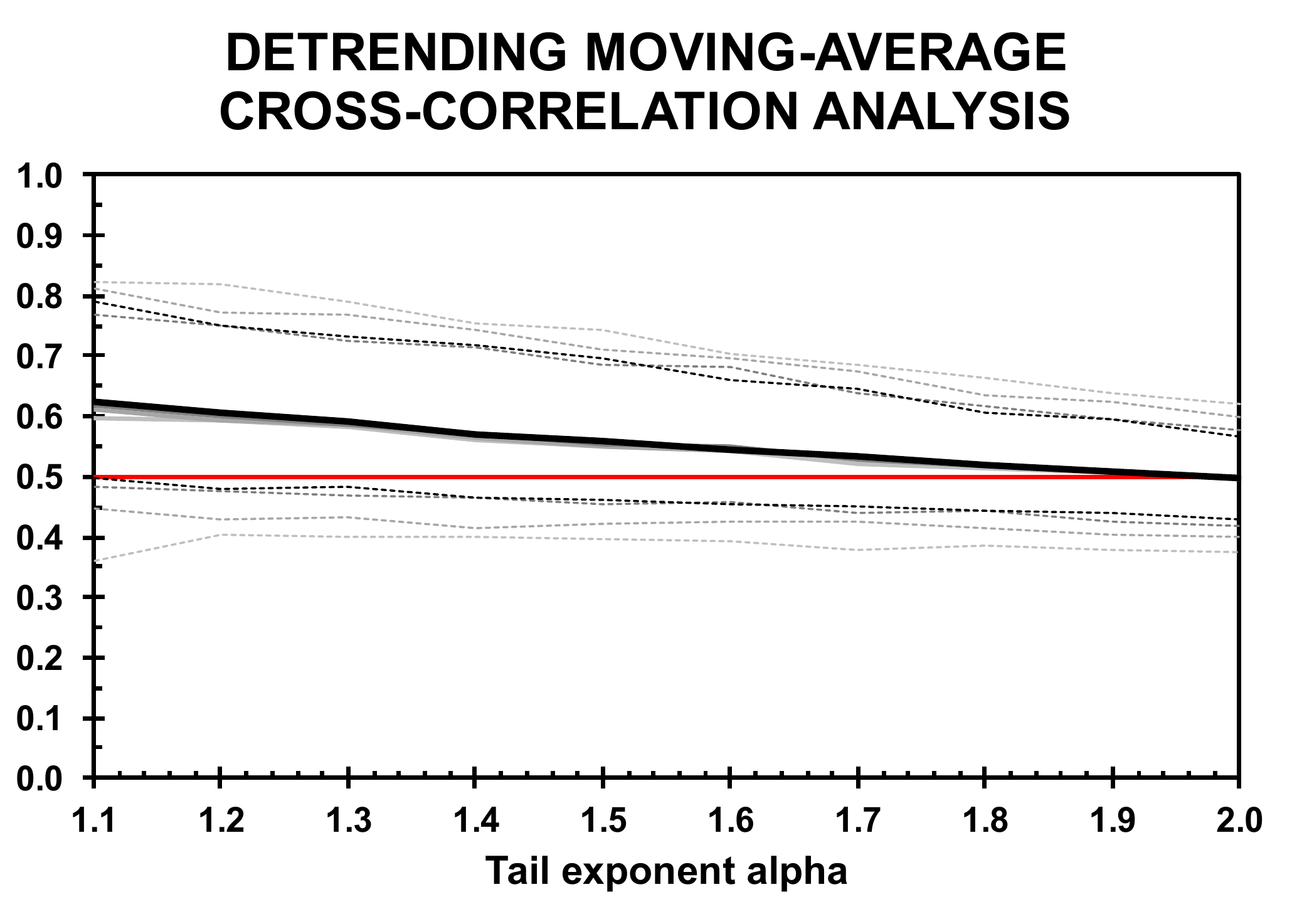}&\includegraphics[width=75mm]{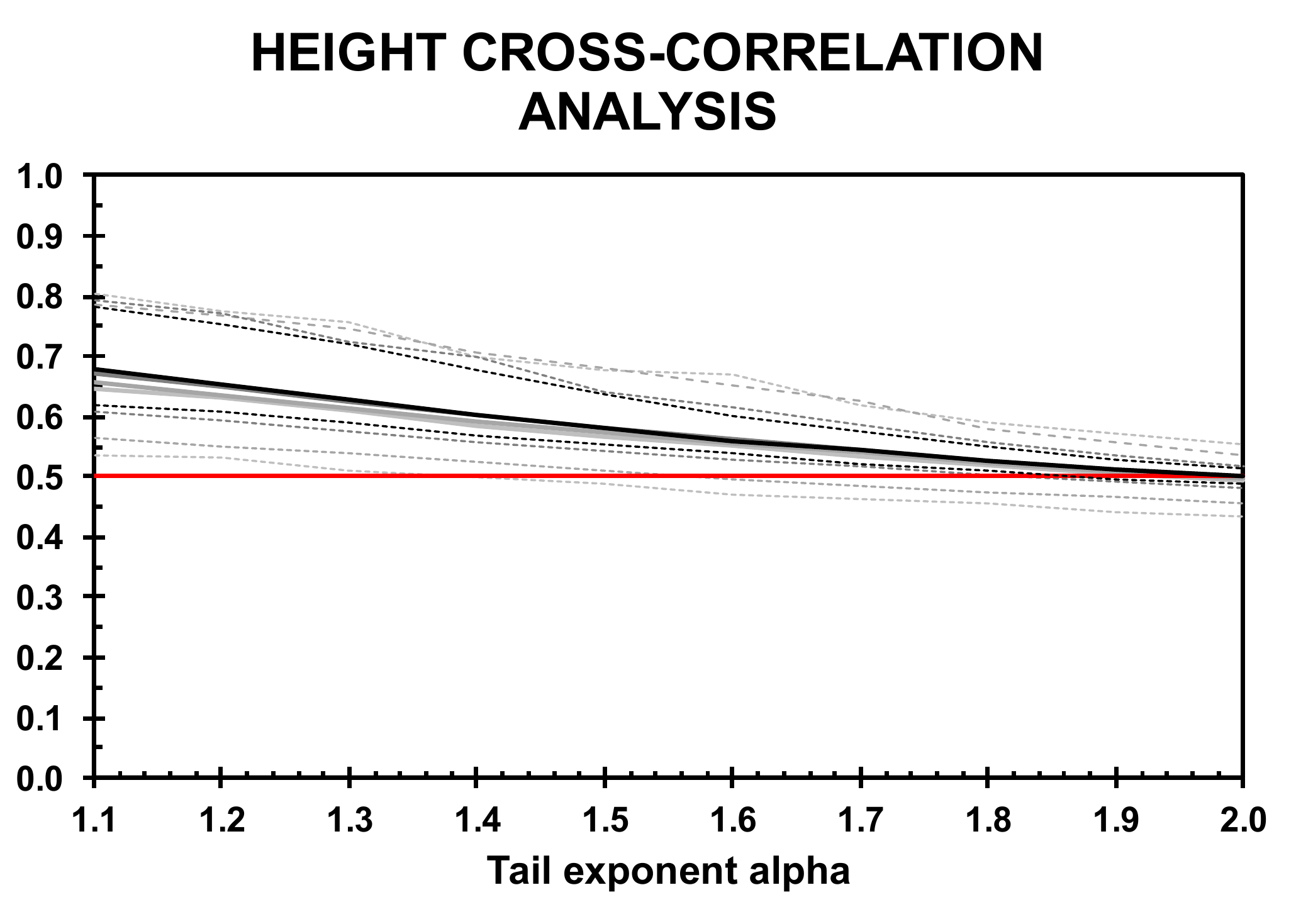}\\
\end{tabular}
\caption{\textbf{Mean values of the estimators for different tail exponents (Setting I).} \footnotesize{
The solid lines represent the average values of 1000 simulations for the given setting. The dashed lines show the 95\% confidence intervals, i.e. the 2.5th and the 97.5th quantiles. The $x$-axis gives values of the tail exponent of series $\{y_t\}$ simulated as the $\alpha$-stable distributed process with the characteristic exponent $\alpha$, i.e. the more to the left of the axis the heavier the tails of the underlying process. The $\{x_t\}$ process is the standard Gaussian noise. Both processes are standardized to have a unit variance. The shades of grey stand for the different time series lengths, $T=500,1000,5000,10000$, the darker the color the longer the series. The red line stands for the theoretical value of $H_{xy}=0.5$. On the one hand, the frequency domain estimators are unbiased whereas the time domain estimators are biased under heavy tails. On the other hand, the latter ones have narrower confidence intervals (especially for HXA). This is further studied in Fig. \ref{SD} and \ref{MSE}.
}\label{Bias}}
\end{center}
\end{figure}

\begin{figure}[!htbp]
\begin{center}
\begin{tabular}{cc}
\includegraphics[width=75mm]{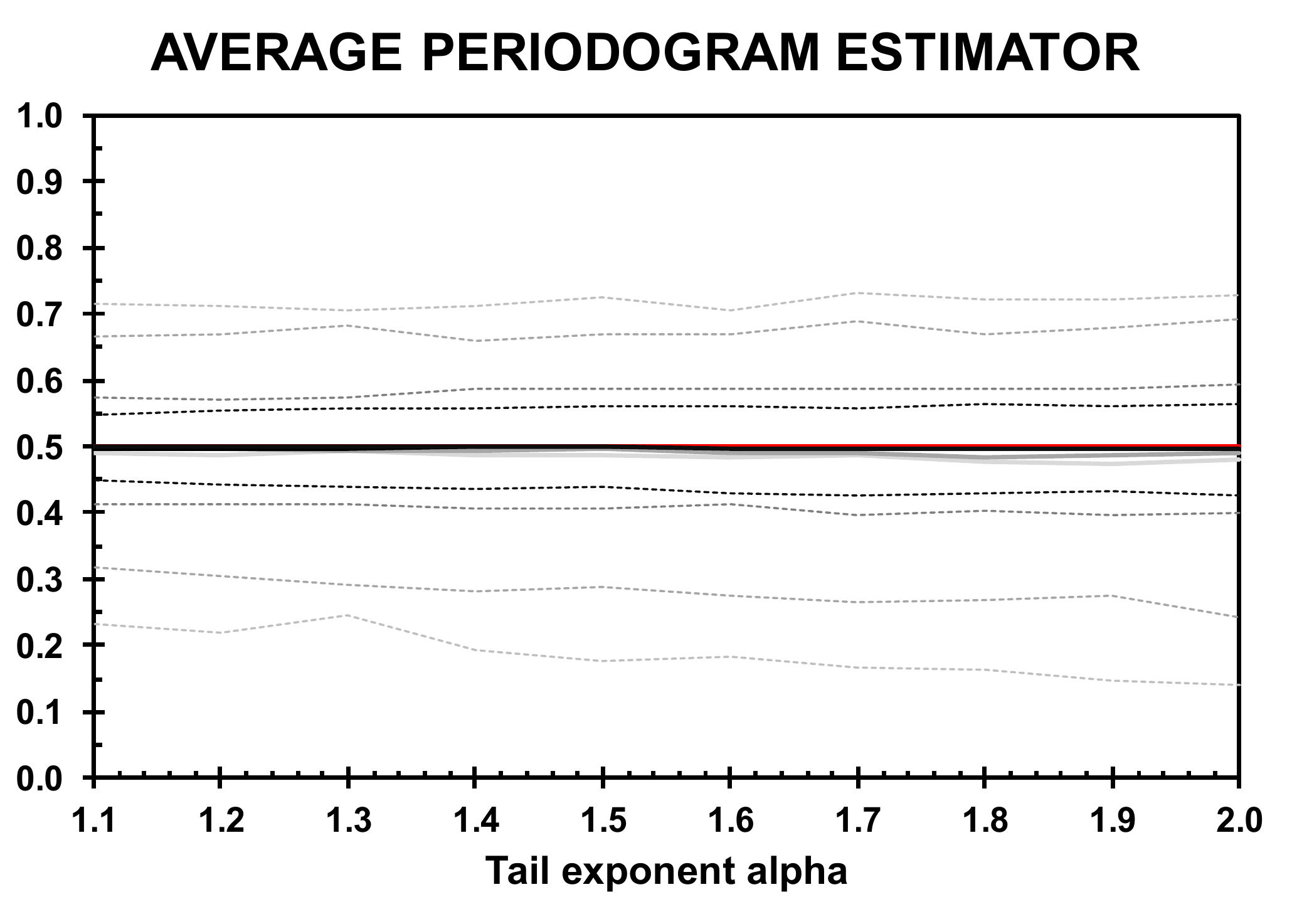}&\includegraphics[width=75mm]{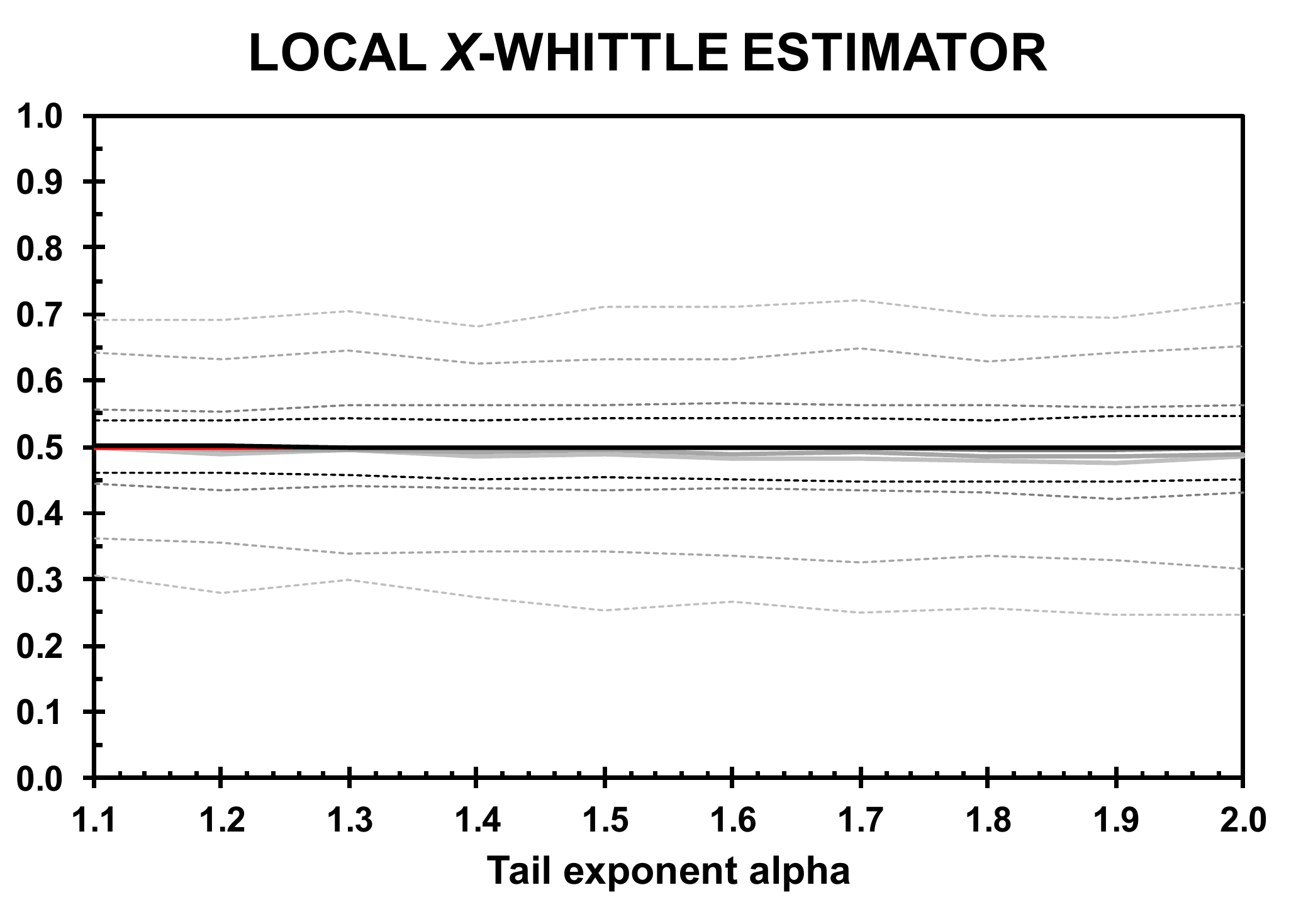}\\
\includegraphics[width=75mm]{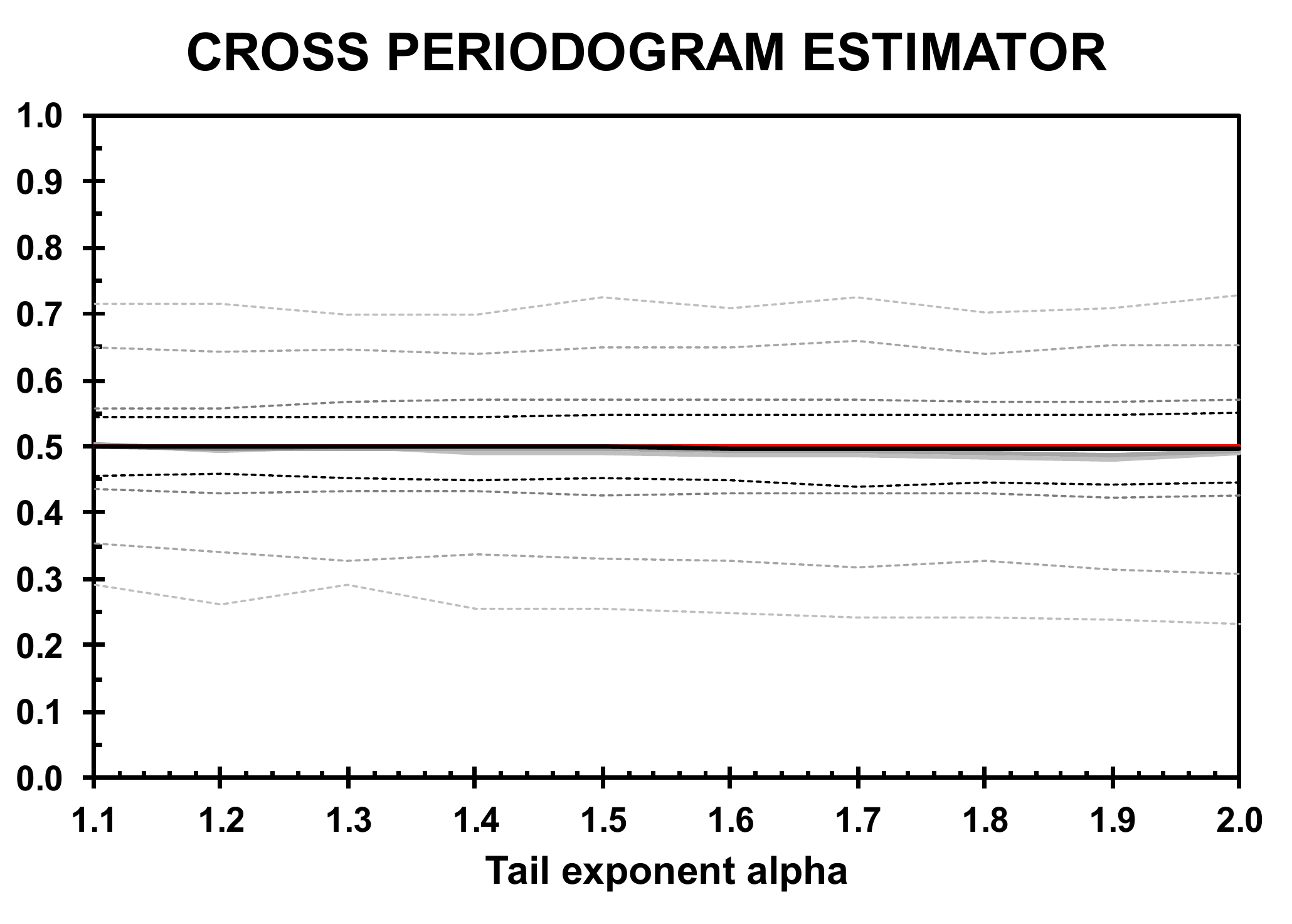}&\includegraphics[width=75mm]{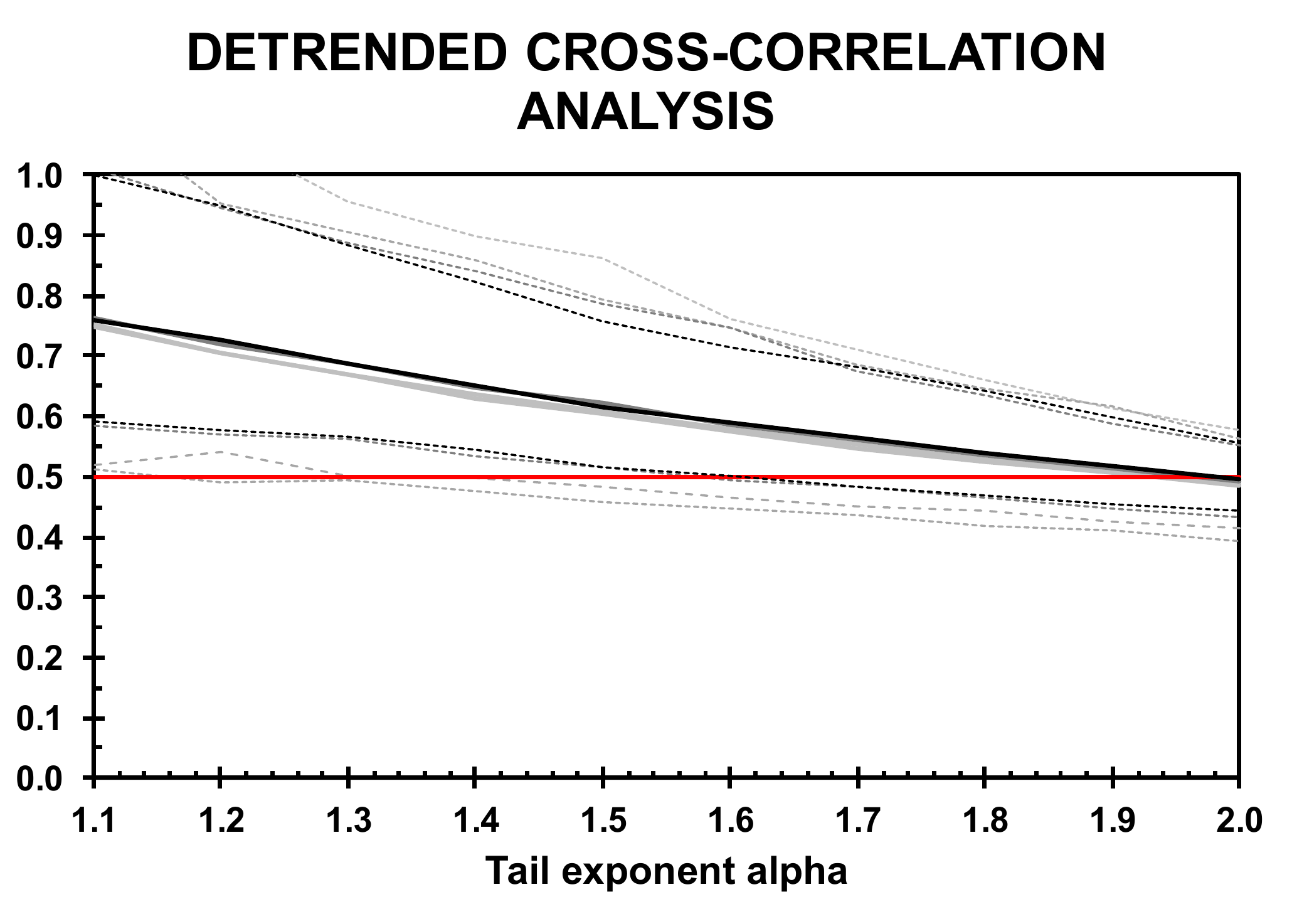}\\
\includegraphics[width=75mm]{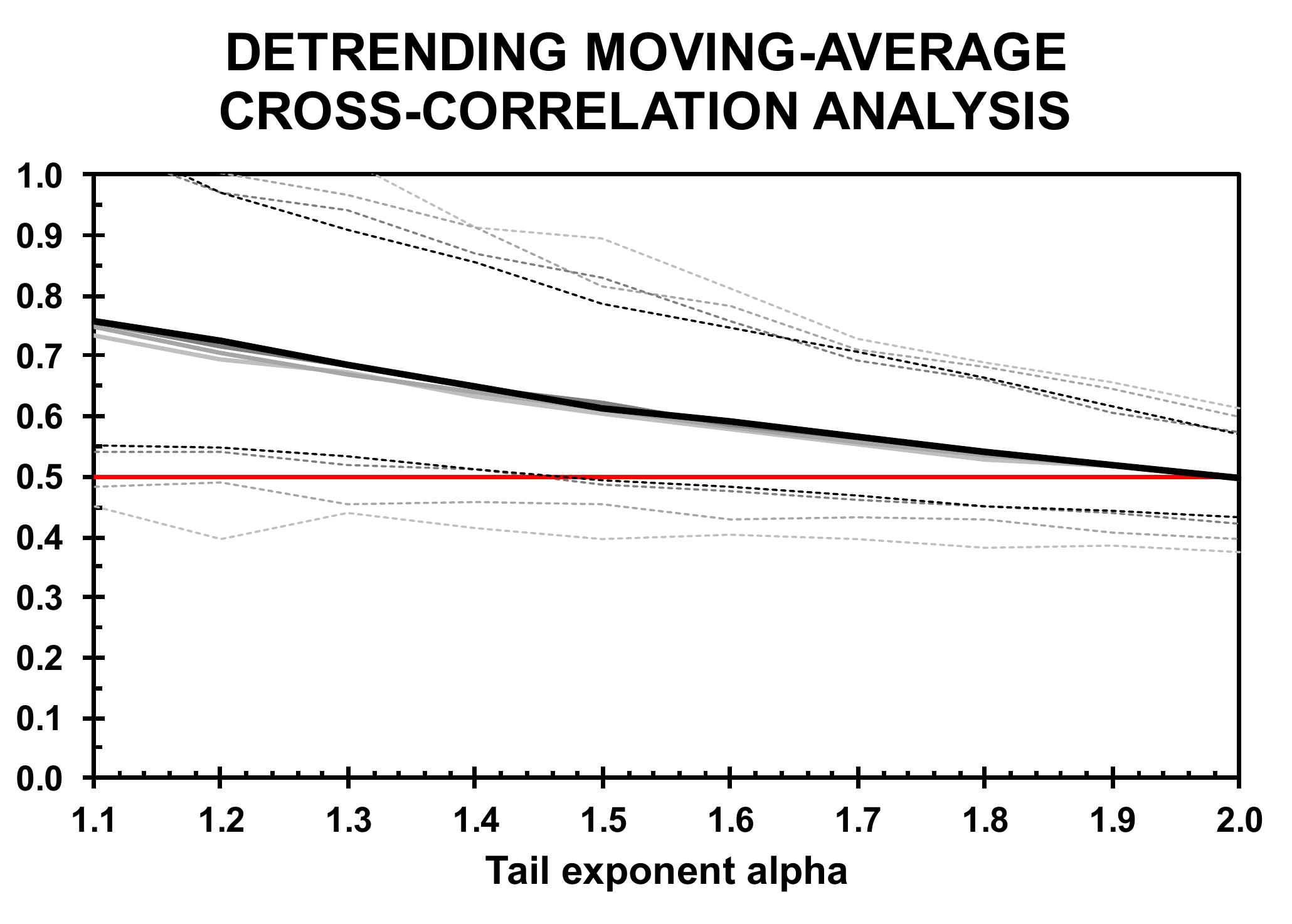}&\includegraphics[width=75mm]{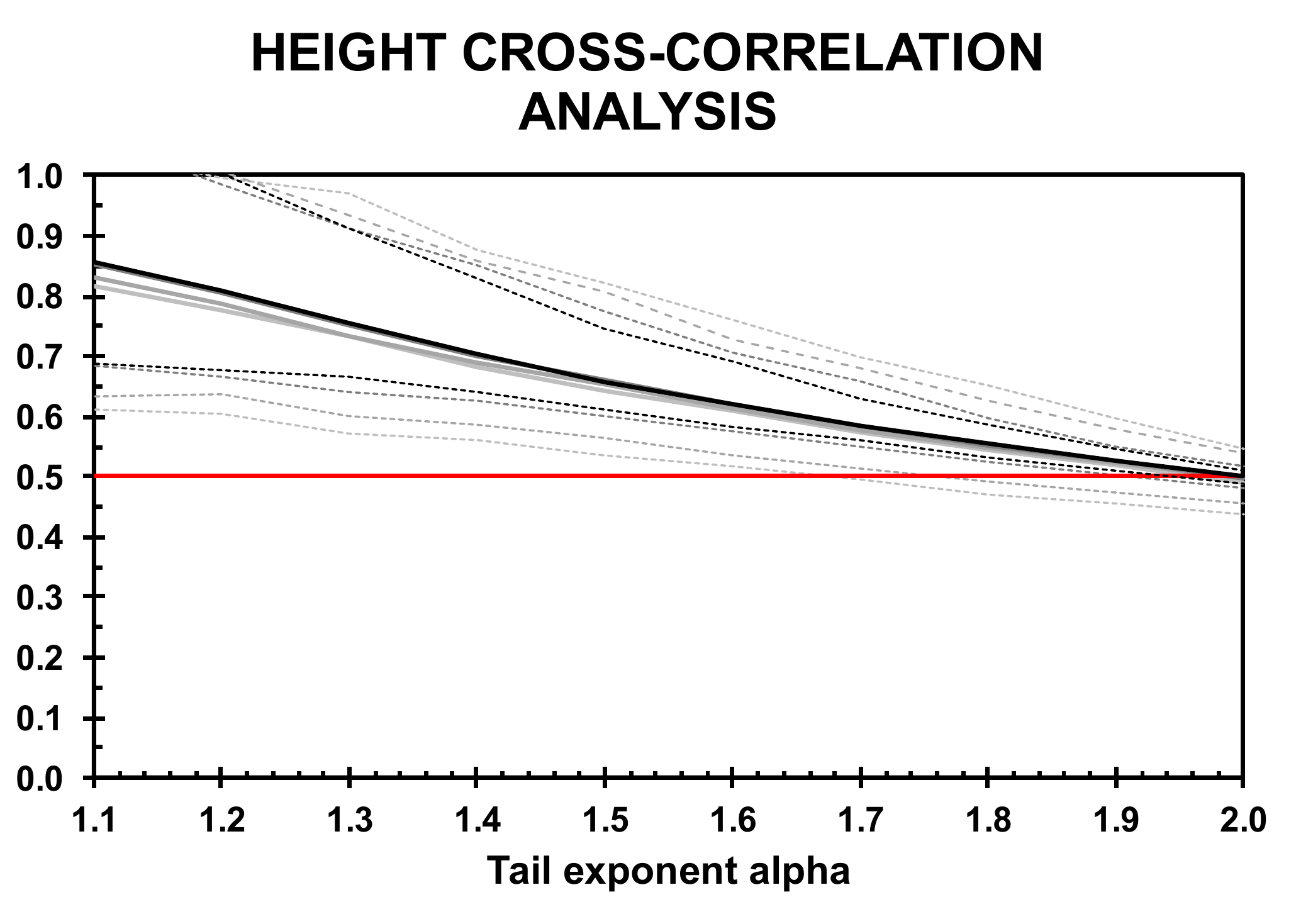}\\
\end{tabular}
\caption{\textbf{Mean values of the estimators for different tail exponents (Setting II).} \footnotesize{
The solid lines represent the average values of 1000 simulations for the given setting. The dashed lines show the 95\% confidence intervals, i.e. the 2.5th and the 97.5th quantiles. The $x$-axis gives values of the tail exponent of series $\{x_t\}$ and $\{y_t\}$, which are both simulated as the $\alpha$-stable distributed process with the characteristic exponent $\alpha$, i.e. the more to the left of the axis the heavier the tails of the underlying process. The shades of grey stand for the different time series lengths, $T=500,1000,5000,10000$, the darker the color the longer the series. The red line stands for the theoretical value of $H_{xy}=0.5$. The results are qualitatively very similar to Setting I illustrated in Fig. \ref{Bias} with differences being more pronounced. This is further studied in Fig. \ref{SD_both} and \ref{MSE_both}.
}\label{Bias_both}}
\end{center}
\end{figure}

We start the results description by examination of bias of the estimators. In Figs. \ref{Bias} and  \ref{Bias_both}, we present the mean values (full line), and the 2.5th and 97.5th quantiles (dashed line) based on 1000 simulations for each setting. The darker the line, the higher the time series length ($T=500,1000,5000,10000$). The red line stands for the expected value of $H_{xy}$ for the given setting, i.e. $H_{xy}=0.5$ for all specifications. Fig. \ref{Bias} shows the results for the setting with one Gaussian and one $\alpha$-stable process, Fig. \ref{Bias_both} represents two $\alpha$-stable processes with same $\alpha$. The outcome is very straightforward. The frequency domain estimators are unbiased for all settings of $\alpha$ and even the quantiles are very stable for different characteristic exponents. These estimators are thus practically untouched by heavy tails of the underlying series. The confidence bands become narrower with the increasing time series length, which is desirable. This is true for both settings. The situation differs for the time domain estimators. For low levels of $\alpha$, i.e. for heavy tails, the estimators are biased upwards. This could result in spuriously reported cross-persistence. The bias decreases with an increasing $\alpha$ and it vanishes for $\alpha=2$ (normal tails). Interestingly, the confidence bands of the estimates get narrower with lighter tails and even for some cases of heavy tails (mainly for shorter series), the bands are narrower than the ones of the frequency domain estimators. The situation is even more pronounces for the setting with two heavy-tailed distributions. For very heavy tails, there are situations when the confidence bands climb above the reasonable value of $H_{xy}=1$.

\begin{figure}[!htbp]
\begin{center}
\begin{tabular}{cc}
\includegraphics[width=75mm]{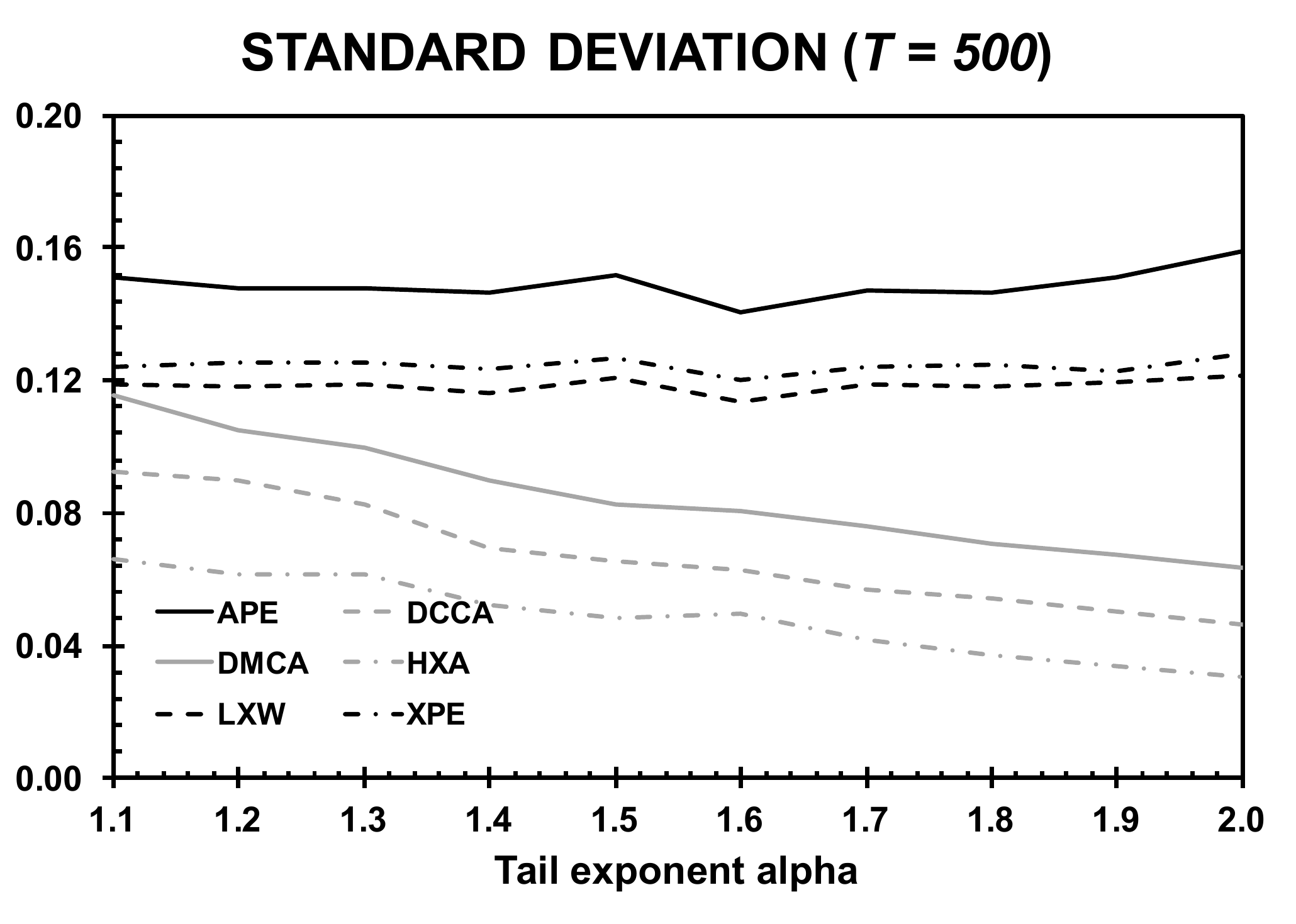}&\includegraphics[width=75mm]{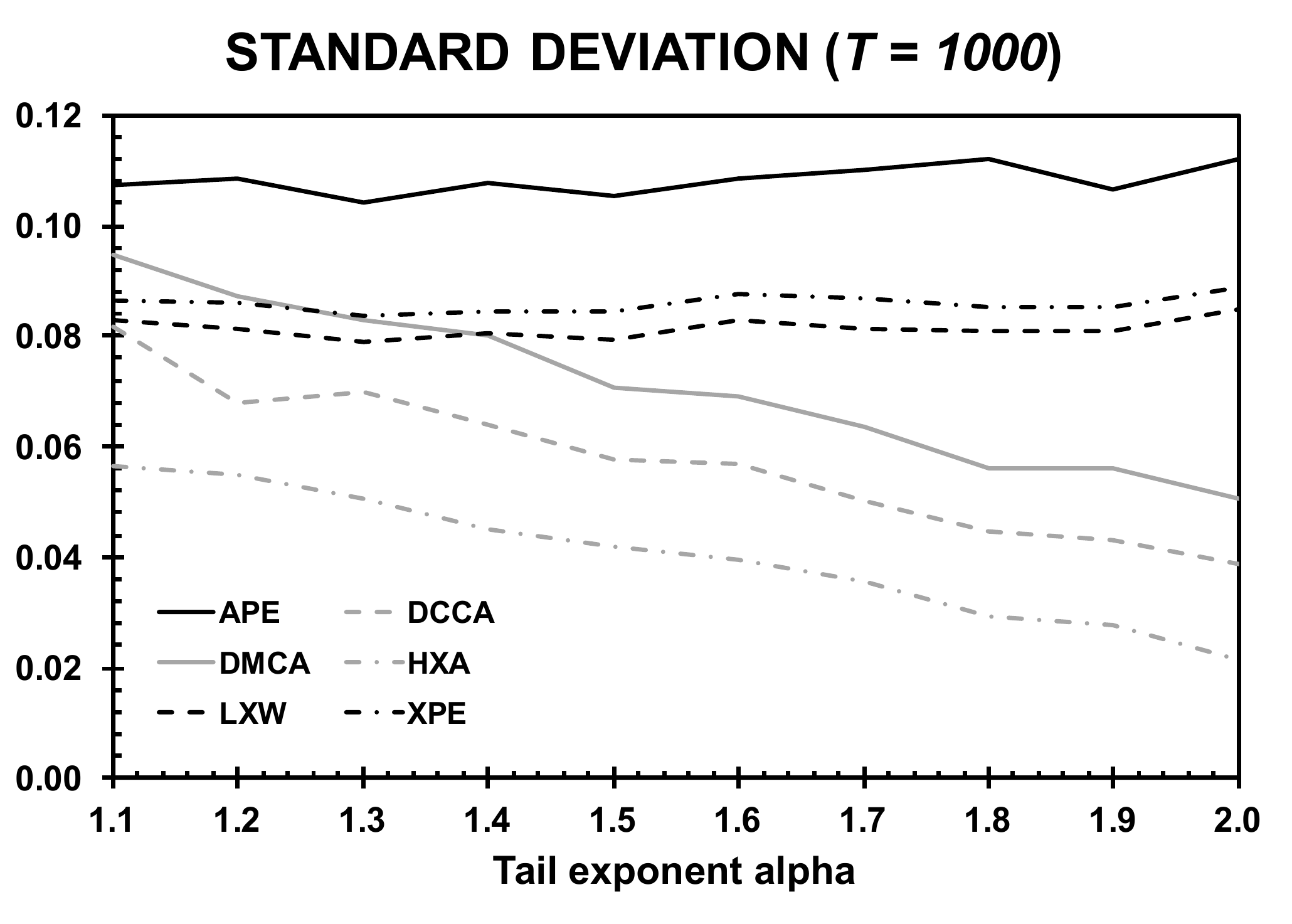}\\
\includegraphics[width=75mm]{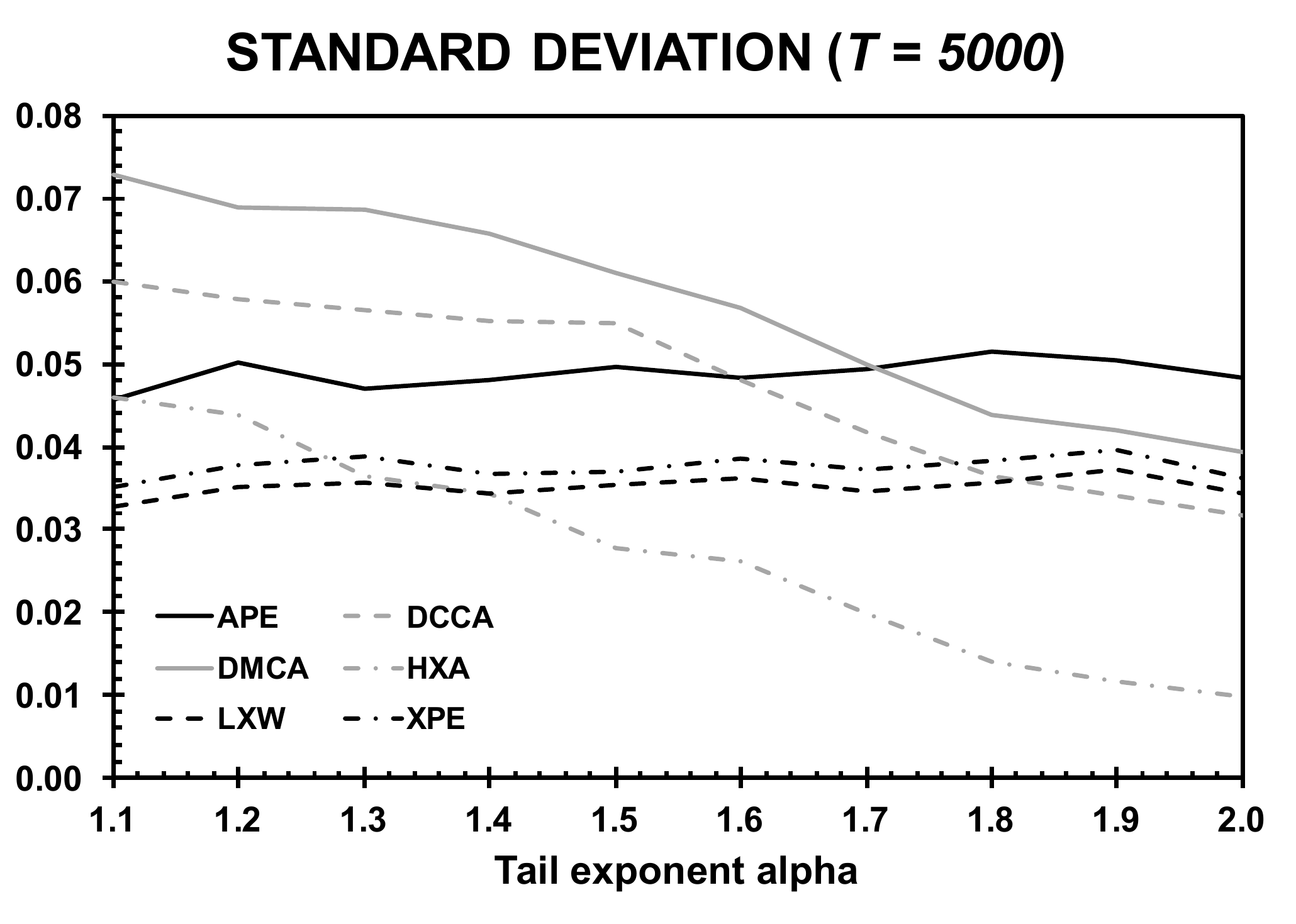}&\includegraphics[width=75mm]{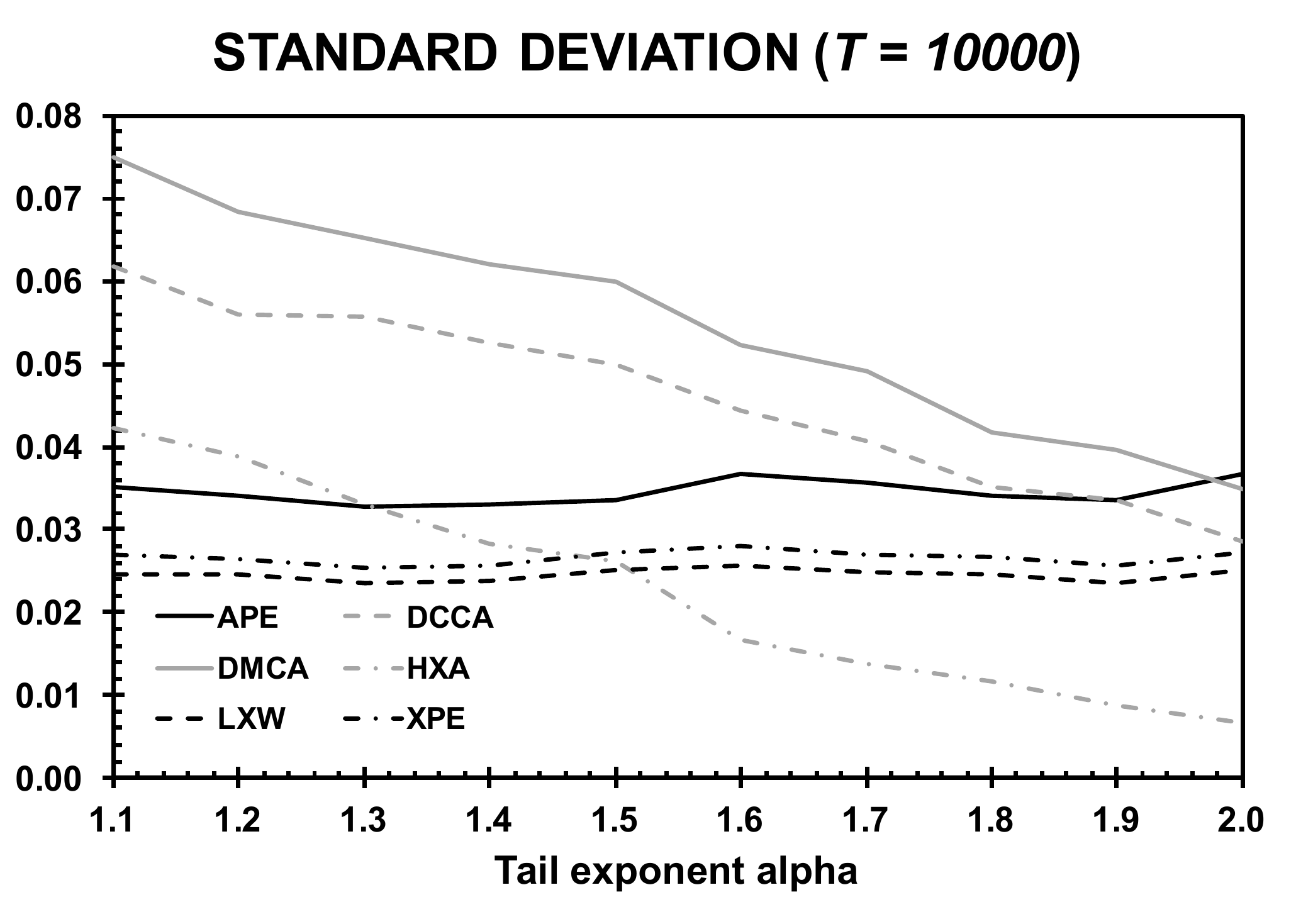}\\
\end{tabular}
\caption{\textbf{Standard deviation of the estimators for different tail exponents (Setting I).} \footnotesize{
The figures show the dependence of estimators' standard deviation on the tail exponent as well as the time series length. The variance is very stable for the frequency domain estimators. For the time domain estimators, the variance depends on the tail exponent strongly. For the shorter series, i.e. $T=500,1000$, the variance of the latter group is much lower compared to the former group. The differences shrink considerably for the longer series, i.e. $T=5000,10000$, where the results become mixed and more in favor of the frequency domain estimators.
}\label{SD}}
\end{center}
\end{figure}

\begin{figure}[!htbp]
\begin{center}
\begin{tabular}{cc}
\includegraphics[width=75mm]{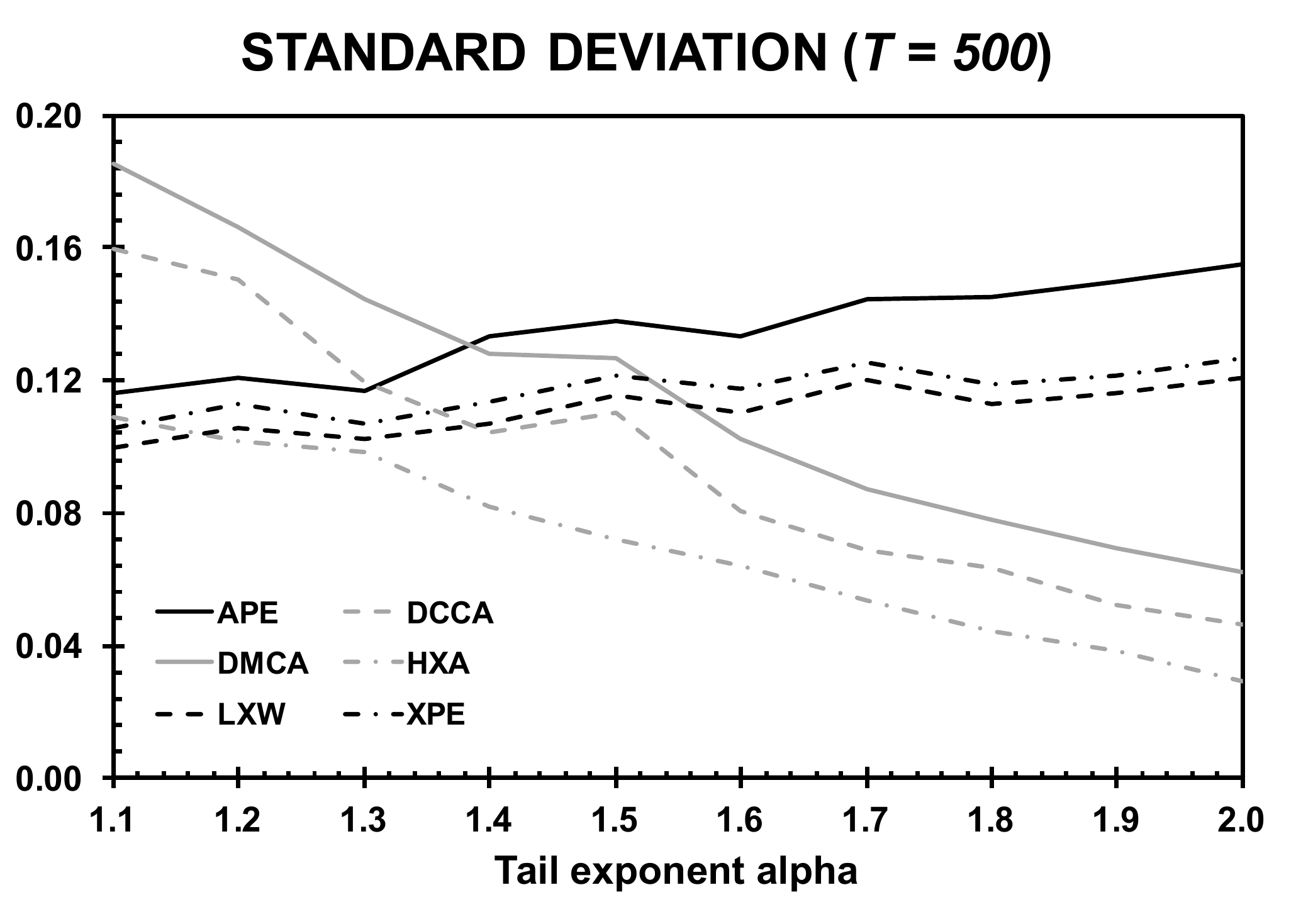}&\includegraphics[width=75mm]{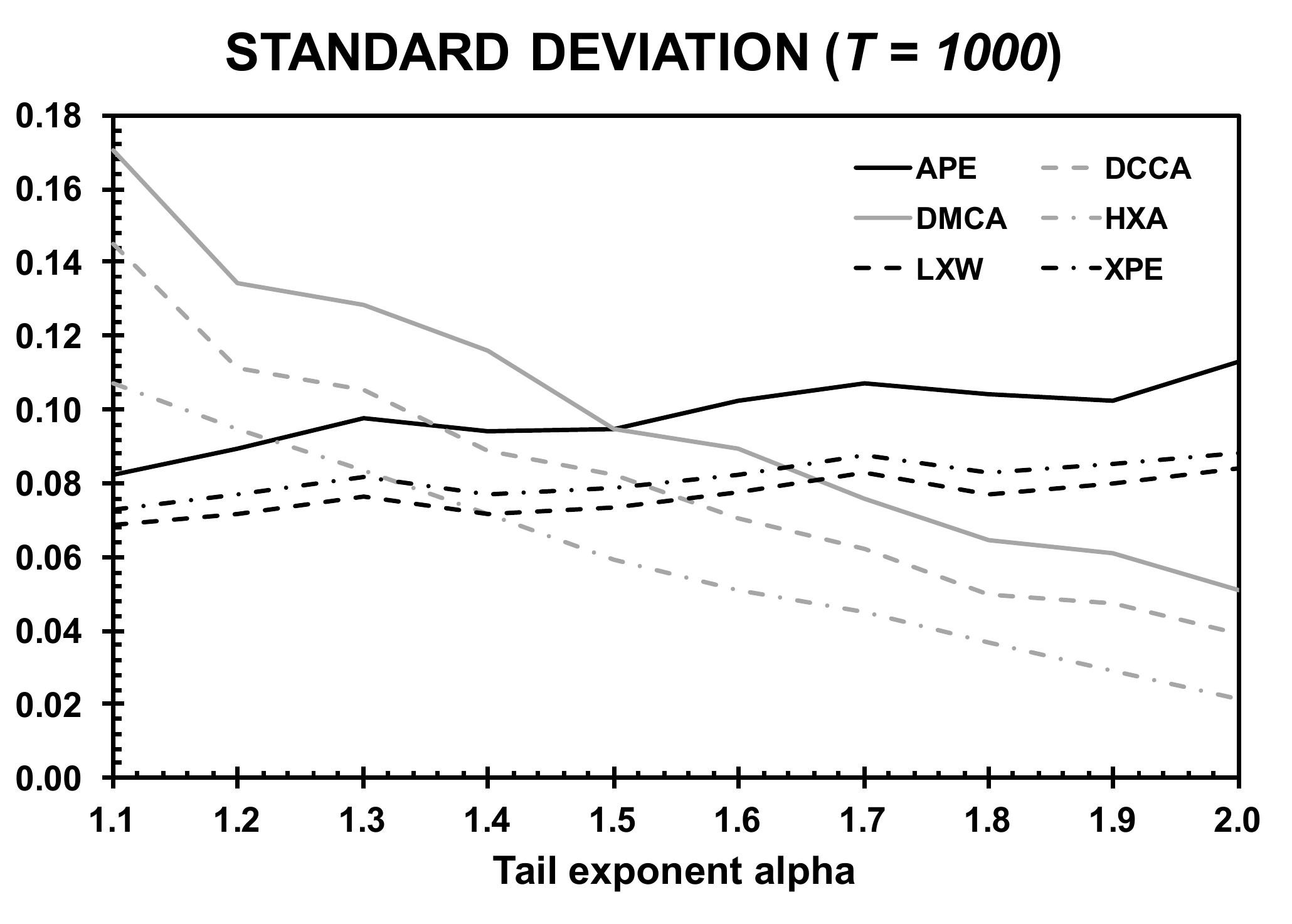}\\
\includegraphics[width=75mm]{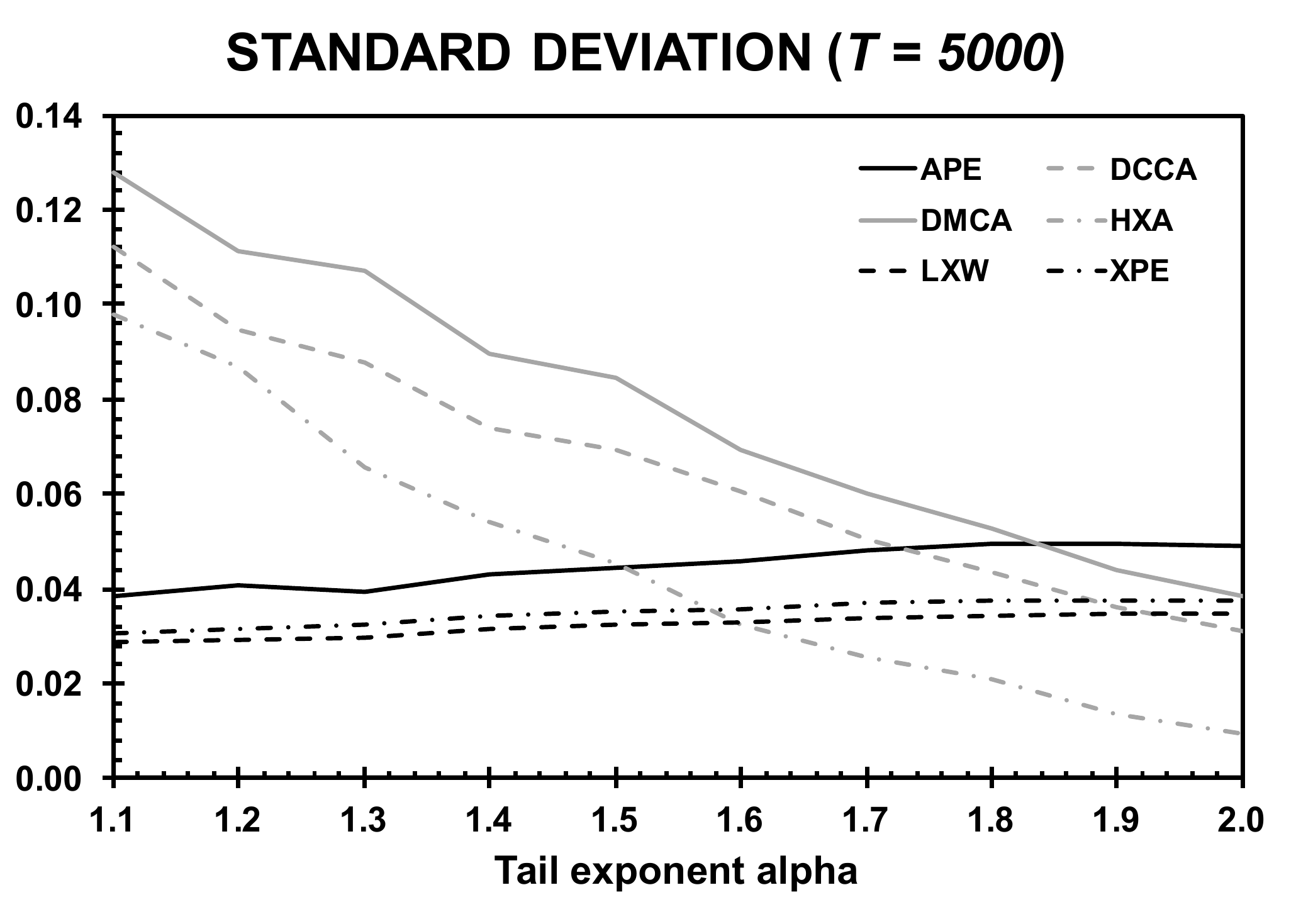}&\includegraphics[width=75mm]{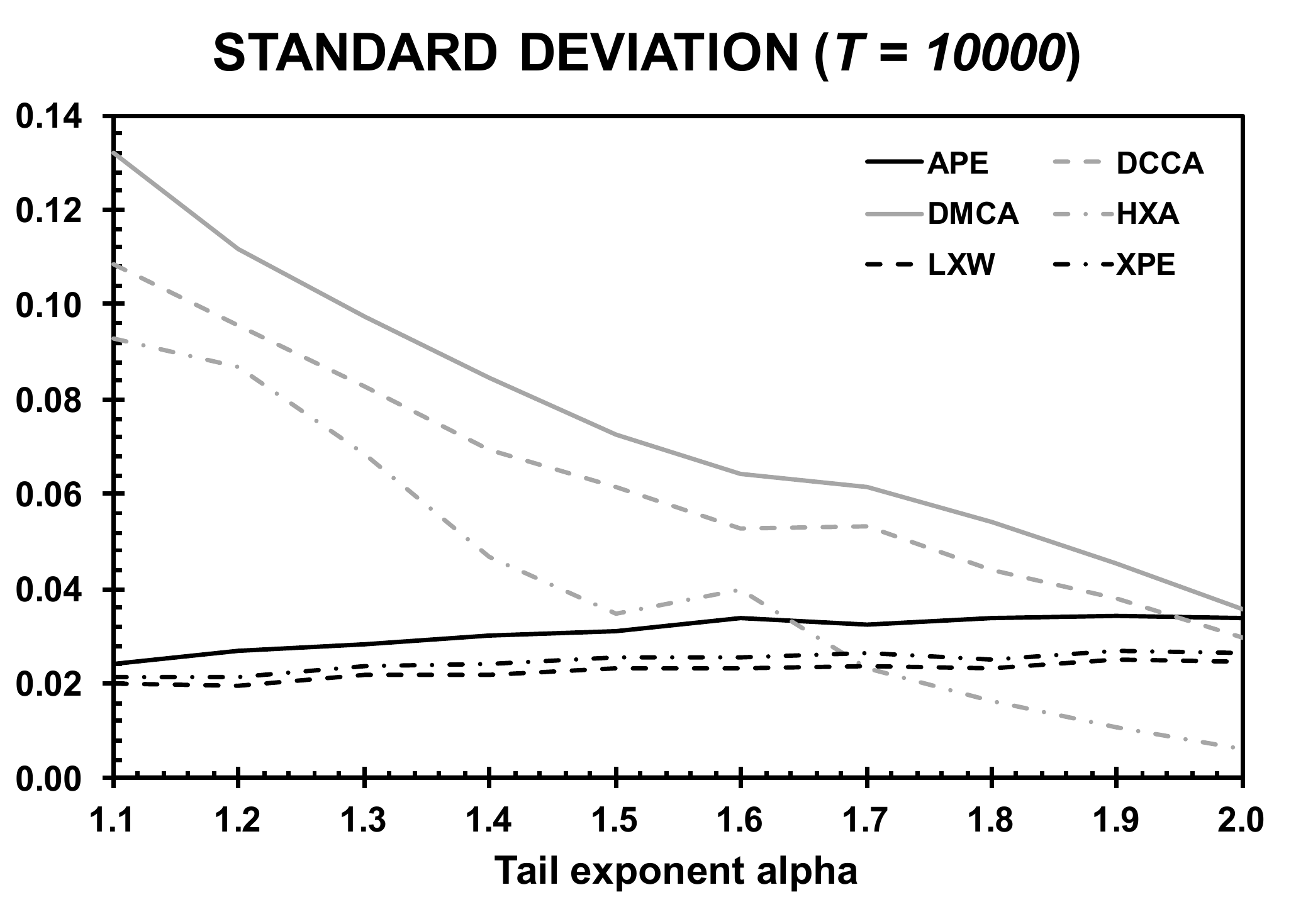}\\
\end{tabular}
\caption{\textbf{Standard deviation of the estimators for different tail exponents (Setting II).} \footnotesize{
The figures show the dependence of estimators' standard deviation on the tail exponent as well as the time series length. The variance is again very stable for the frequency domain estimators. It even slightly increasing with lighter tails. For the time domain estimators, the variance depends on the tail exponent strongly. For the shorter series, i.e. $T=500,1000$, and lighter tails, i.e. approximately $\alpha>1.5$, the variance of the latter group is much lower compared to the former group. The differences shrink considerably for the longer series, i.e. $T=5000,10000$, where the frequency domain estimators strongly dominate for the most levels of tails.
}\label{SD_both}}
\end{center}
\end{figure}

To further examine this discrepancy, Figs. \ref{SD} and \ref{SD_both} report the estimator standard deviation dependence on parameter $\alpha$ for all investigated estimators. The results obviously reflect the results for confidence bands but these are more transparent here (the figures are split based on the specific time series lengths for an easier comparison between estimators). In fact, standard deviation of the frequency domain estimators is practically independent of $\alpha$. The highest standard deviation is reported for APE followed by XPE and LXW with quite similar results. For short time series with $T=500$ and $T=1000$, the time domain estimators possess lower variance for practically all levels of $\alpha$. The lowest variance is found for HXA followed by DCCA and DMCA, respectively. The evidence gets more mixed for $T=5000$. And for $T=10000$, DCCA and DMCA estimators are dominated by the frequency domain estimators for all levels of $\alpha$ but the Gaussian case. The HXA estimator outperforms even the frequency domain estimators for $\alpha>1.5$. These implications are valid for the case of one Gaussian and one $\alpha$-stable distribution in the simulations (Fig. \ref{SD}). However, the situation gets more clearcut for the case of two $\alpha$-stable distributions (Fig. \ref{SD_both}). Standard deviation of the frequency domain estimators is again reasonably stable across different $\alpha$. Interestingly, their standard deviation even slightly increases with lighter tails (increasing $\alpha$). The general level of the time domain estimators standard deviations increases markedly (compare Figs. \ref{SD} and \ref{SD_both}). Now even for short series, the tails need to be quite light ($\alpha>1.5$ for $T=500$ and $\alpha>1.7$ for $T=1000$) for the time domain estimators to clearly dominate. For the longer series with $T=5000$ and $T=10000$, only HXA competes with the frequency domain estimators and it outperforms them only for $\alpha>1.6$ and $\alpha>1.7$ for $T=5000$ and $T=10000$, respectively. For the frequency domain estimators, it again holds that XPE and LXW are the best performing estimators while LXW has a slight edge over XPE.

A closer look at both Figs. \ref{SD} and \ref{SD_both} uncover a clear difference between the behavior of standard deviations with respect to time series length for both groups of estimators. For the frequency domain estimators, we can easily observe that their standard deviation decays approximately at a rate of $1/\sqrt{T}$ (or $1/T$ for variance). However, the decay is much slower for the time domain estimators (at approximately $T^{-0.15}$ for standard deviation). This is a reflection of the univariate frequency domain estimators being asymptotically normal and efficient \citep{Robinson1994,Robinson1995a} whereas no such quality is known for the time domain estimators.

\begin{figure}[!htbp]
\begin{center}
\begin{tabular}{cc}
\includegraphics[width=75mm]{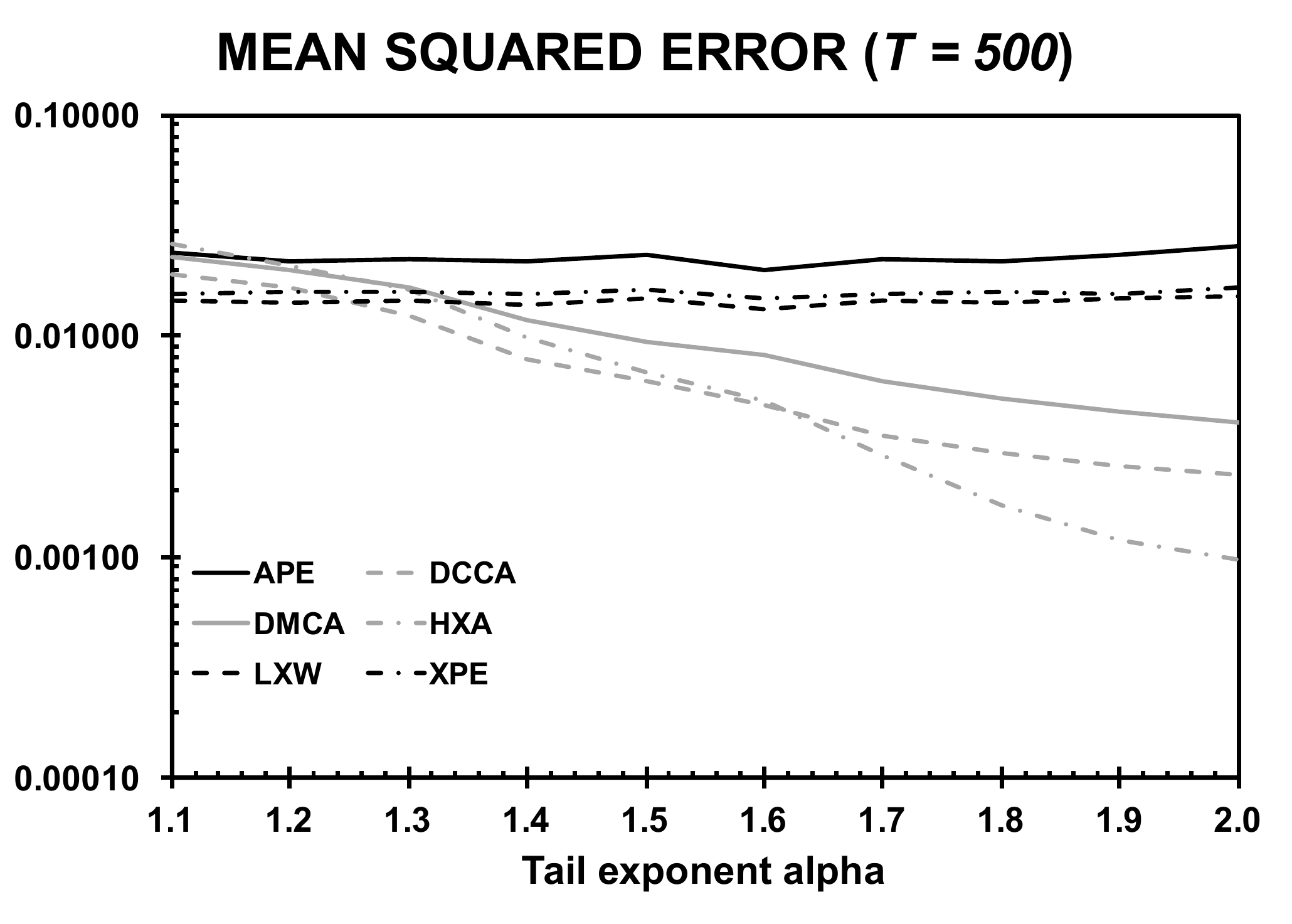}&\includegraphics[width=75mm]{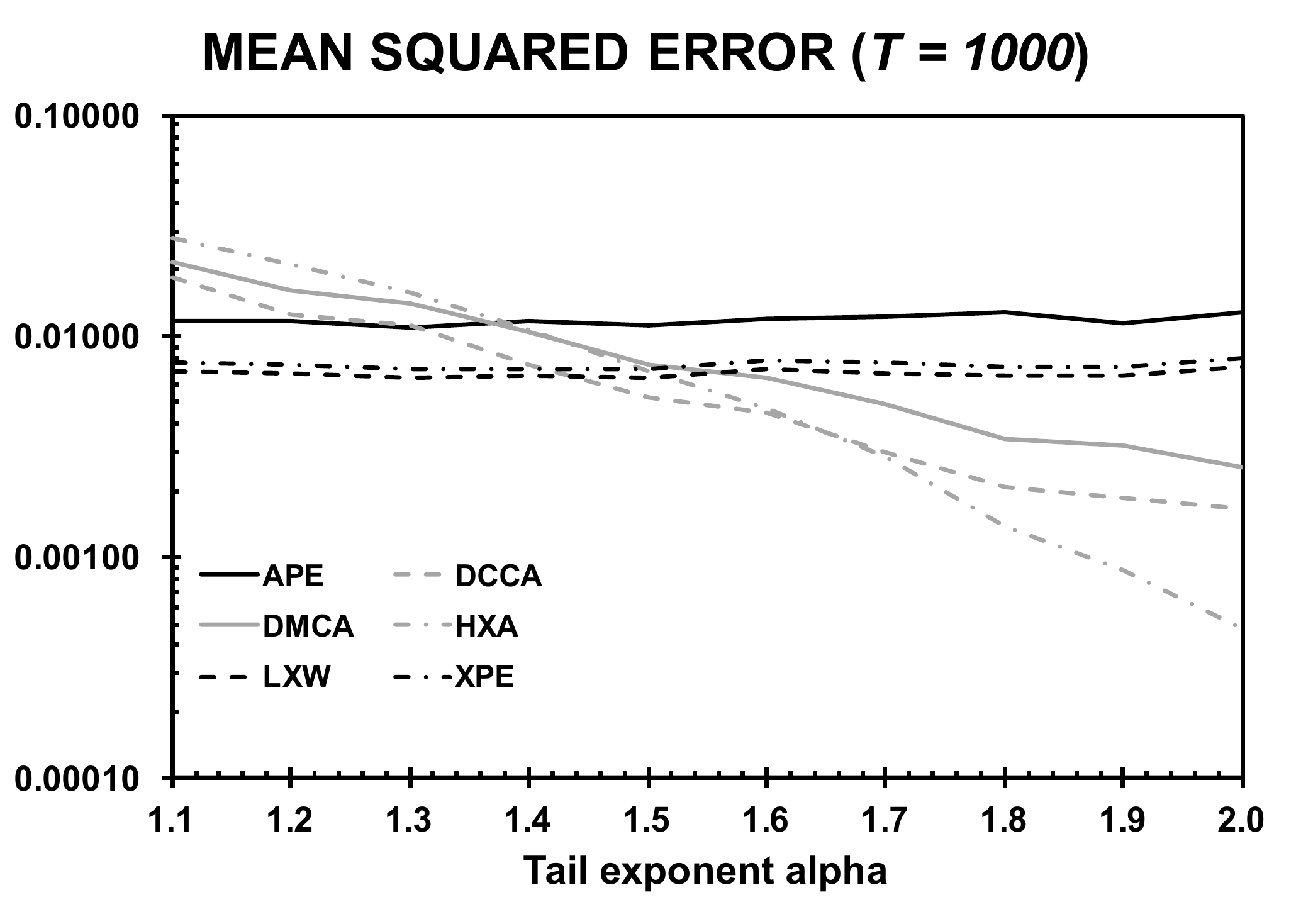}\\
\includegraphics[width=75mm]{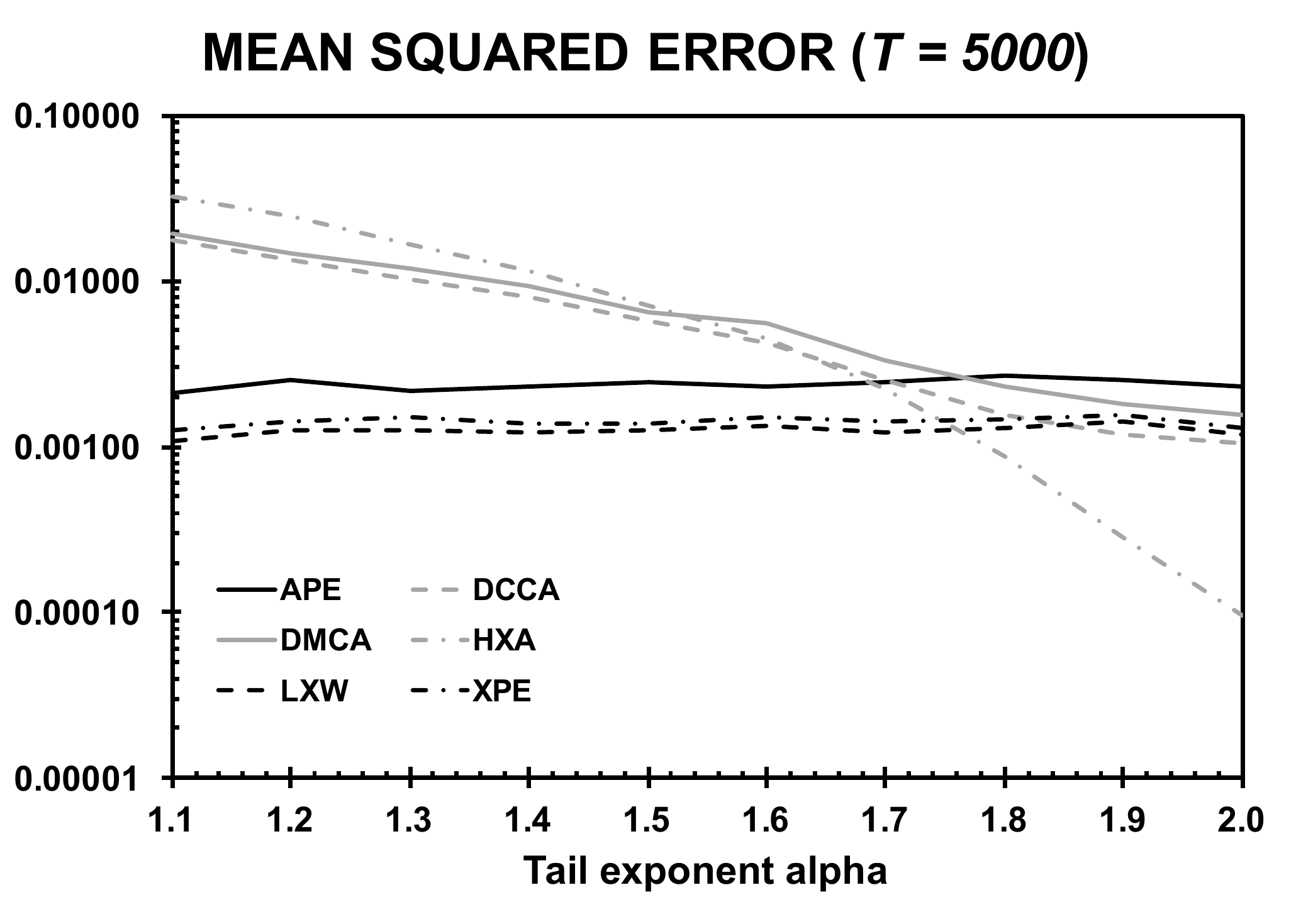}&\includegraphics[width=75mm]{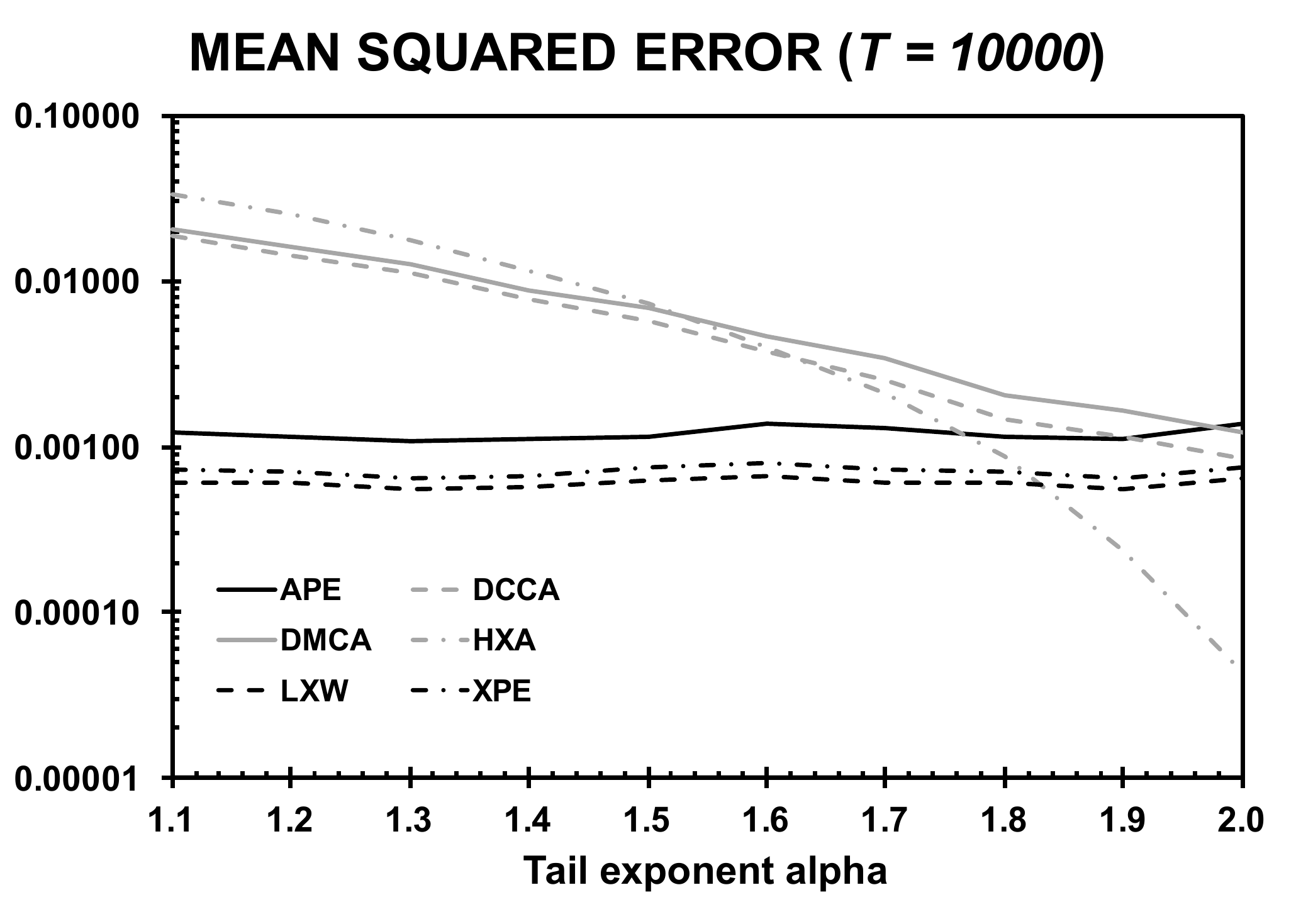}\\
\end{tabular}
\caption{\textbf{Mean squared error of the estimators for different tail exponents (Setting I).} \footnotesize{
Mean squared error combines the performance of the estimators based on their bias as well as their variance. MSE is stable for the frequency domain estimators but strongly dependent on the tail exponent for the time domain estimators. The discrepancy between the two groups is dependent both on the tail index and on the time series length. For the shorter series, i.e. $T=500,1000$, the latter group outperforms the former for most tail index values. For the longer series, the former group dominates apart from tails close to normality where the HXA method strongly dominates all the other methods.
}\label{MSE}}
\end{center}
\end{figure}

\begin{figure}[!htbp]
\begin{center}
\begin{tabular}{cc}
\includegraphics[width=75mm]{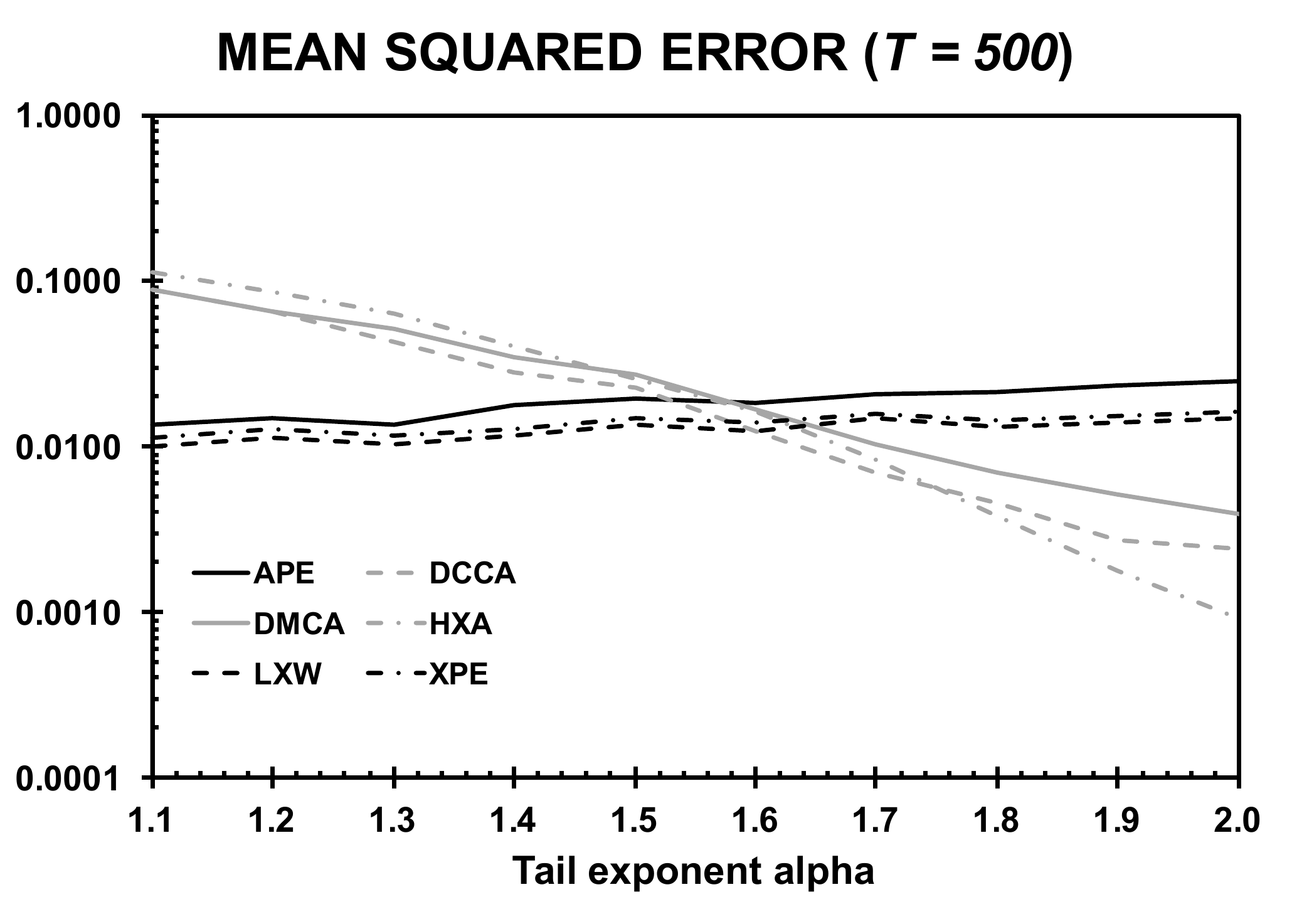}&\includegraphics[width=75mm]{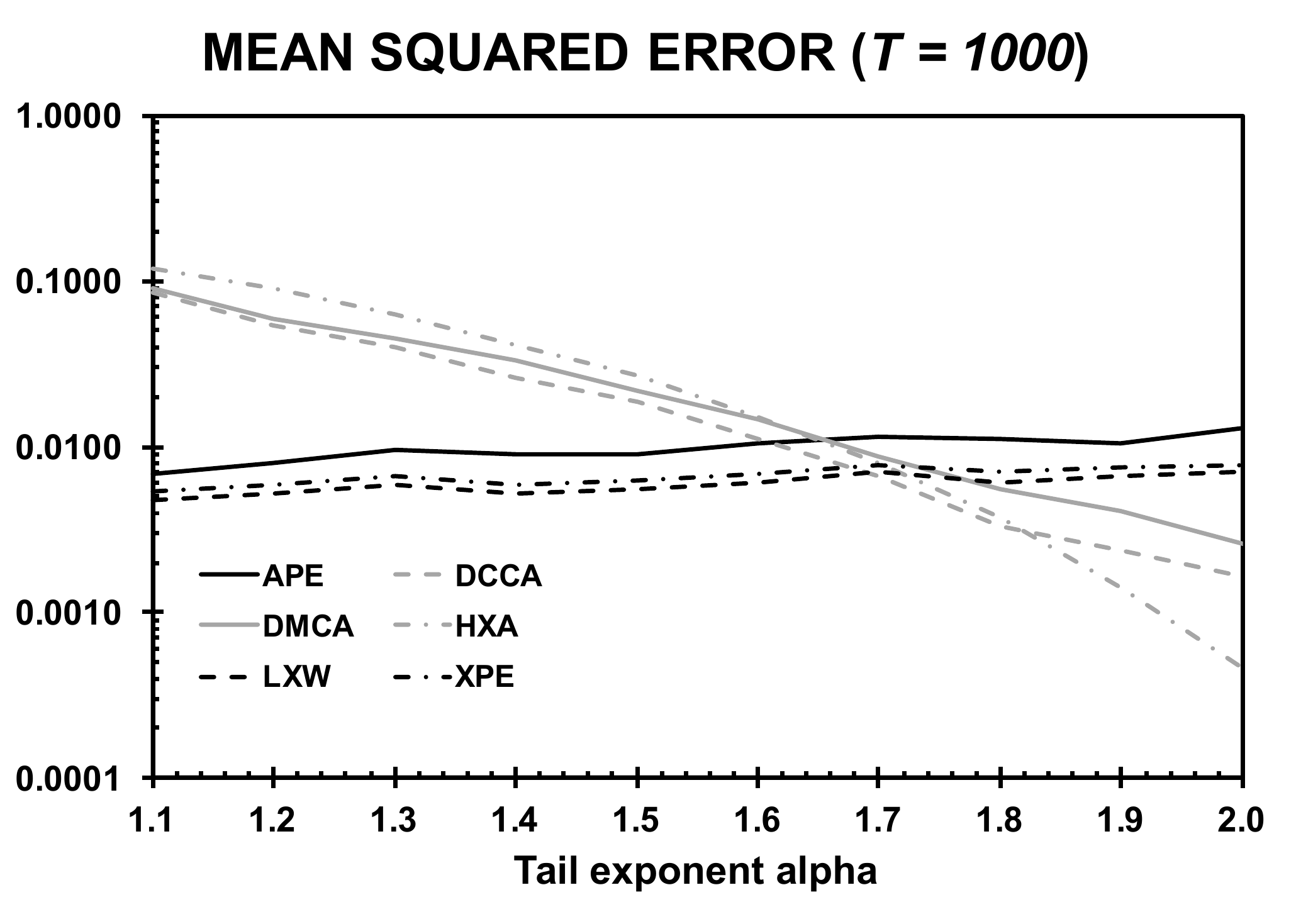}\\
\includegraphics[width=75mm]{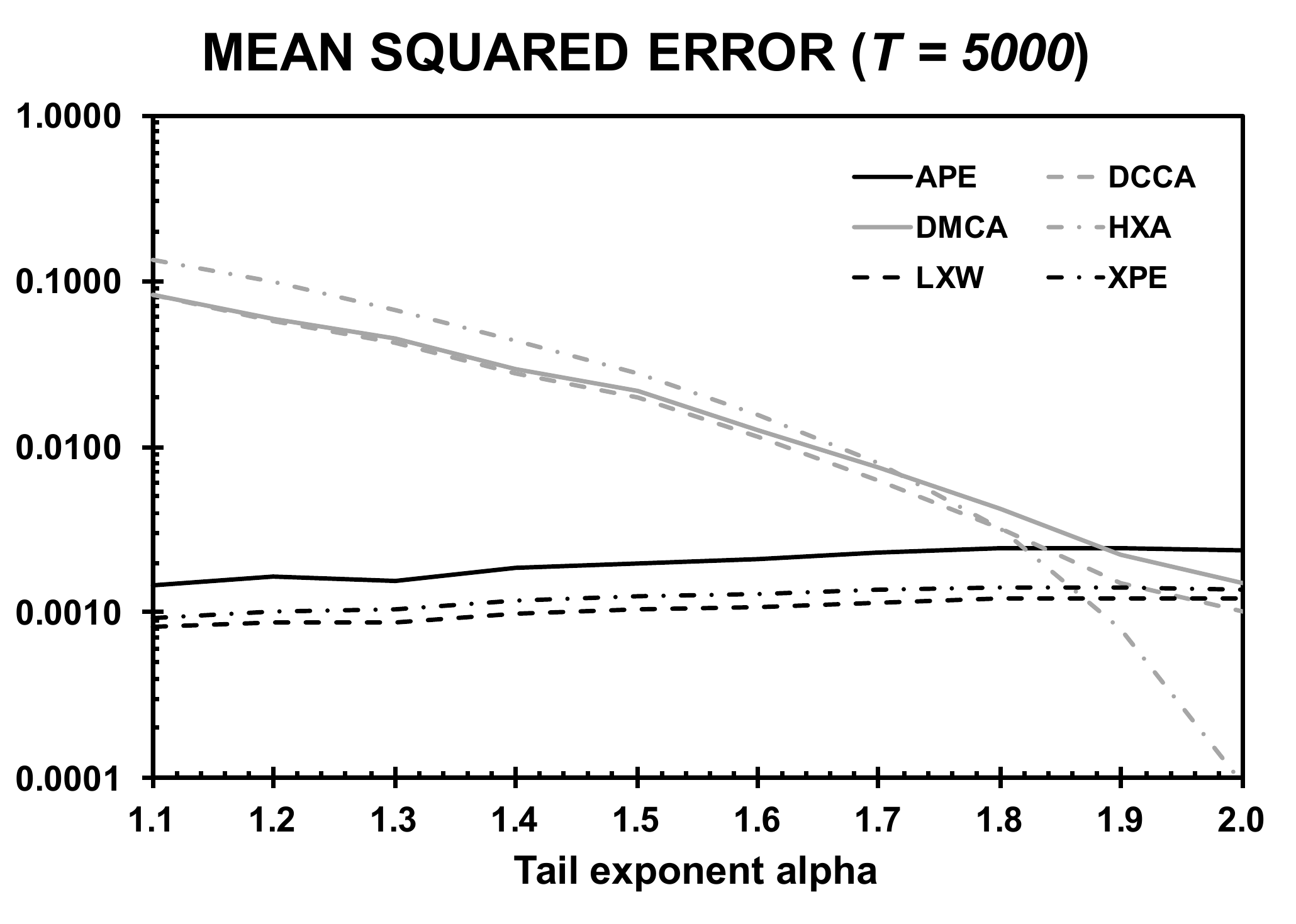}&\includegraphics[width=75mm]{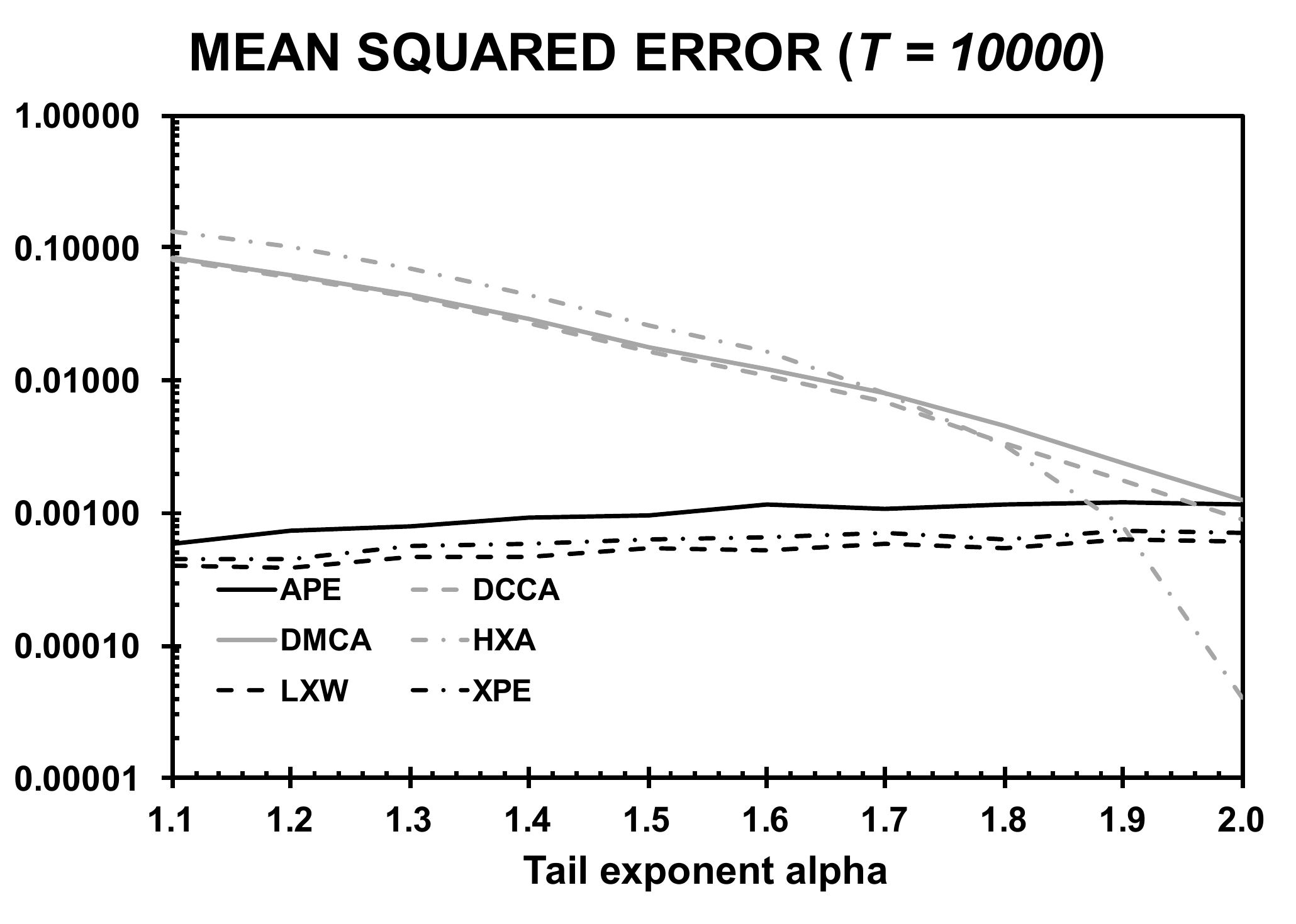}\\
\end{tabular}
\caption{\textbf{Mean squared error of the estimators for different tail exponents (Setting II).} \footnotesize{
Mean squared error combines the performance of the estimators based on their bias as well as their variance. The results are qualitatively very similar to Fig. \ref{MSE}. MSE is again stable for the frequency domain estimators but strongly dependent on the tail exponent for the time domain estimators. The discrepancy between the two groups is dependent both on the tail index and on the time series length. The results are in hand with the standard deviation behavior in Fig. \ref{SD_both}.
}\label{MSE_both}}
\end{center}
\end{figure}

The tradeoff between bias and variance is nicely summed together in the mean squared error (MSE), which comprises both. In Figs. \ref{MSE} and \ref{MSE_both}, we compare MSE for all estimators and different parameter settings. We start with case of one Gaussian and one $\alpha$-stable process. For short time series with $T=500$, the time domain estimators dominate the frequency domain ones for $\alpha>1.3$. The methods are tied for the lower values of $\alpha$. HXA and DCCA come out as winners with HXA having a slight edge over DCCA for light tails ($\alpha > 1.6$). The breaking point increases from $\alpha \approx 1.3$ to $\alpha \approx 1.5$ for $T=1000$ but the results remain qualitatively unchanged. For the longer time series lengths ($T=5000$ and $T=10000$), the frequency domain methods dominate for practically all levels of $\alpha$ apart from light tails with $\alpha \ge 1.8$ where HXA gives more precise estimates in the MSE sense. The other time domain methods do not provide such results for long time series. For the frequency domain estimators, the ranking based on standard deviations translates perfectly to the mean squared error case. Specifically, LXW is the best performing frequency domain estimator tightly followed by XPE. The APE estimator margin is evident. The results are again more pronounced for the case of two $\alpha$-stable distributions. In Fig. \ref{MSE_both}, we can see that the breaking points between better performance of the frequency domain estimators and time domain estimators move up. Specifically, we have the breaking point of $\alpha \approx 1.6$ for $T=500$ and $\alpha \approx 1.7$ for $T=1000$. For the longer series ($T=5000$ and $T=10000$), again only HXA can compete with APE, LXW and XPE and it dominates them only for $\alpha>1.9$. The ranking of the frequency domain estimators remains unchanged compared to the other setting.

The presented results provide clear implications. First, the frequency domain estimators (APE, LXW and XPE) are not affected by heavy tails. Second, the time domain estimators (DCCA, DMCA and HXA) are biased upwards by heavy tails. Third, the time domain estimators have lower variance than the frequency domain ones for shorter time series ($T=500$ and $T=1000$) and lighter tails (above approximately $\alpha=1.5$). Putting the findings together to provide some recommendations for use of the estimators, we suggest using the time domain estimators for short time series and light tails close to the normal ones. Otherwise, we recommend to use the frequency domain estimators, preferably LXW and XPE. Comparison of two different settings of having either one Gaussian and one $\alpha$-stable distributed process or two $\alpha$-stable distributed ones only highlights the sensitivity of the time domain estimators to heavy tails. 

The findings also have important implications for the stream of literature primarily using the time domain estimators. Combined with the results of Barunik \& Kristoufek \cite{Barunik2010} who show that the univariate versions of DCCA, DMCA and HXA (detrended fluctuation analysis, DFA, detrending moving average, DMA, and height-height correlation analysis, HCCA, respectively) are unbiased for various levels of heavy tails, we provide a possible explanation for the frequently reported estimated bivariate Hurst exponent $\widehat{H}_{xy}$ being higher than the average of the separate estimated Hurst exponents $\widehat{H}_x$ and $\widehat{H}_y$, i.e. $\widehat{H}_{xy}>\frac{\widehat{H}_x+\widehat{H}_y}{2}$. As the Hurst exponent estimation is unbiased by heavy tails in the univariate setting, the upward bias in the bivariate setting for heavy tails implies the possibility of $\widehat{H}_{xy}>\frac{\widehat{H}_x+\widehat{H}_y}{2}$. However, the recent study of Kristoufek \cite{Kristoufek2015b} analytically shows that $H_{xy}\le\frac{H_x+H_y}{2}$ for very generally defined processes. In addition, the author argues that one of the possible reasons for reporting $\widehat{H}_{xy}>\frac{\widehat{H}_x+\widehat{H}_y}{2}$ is a finite sample bias. Here we add the bias due to distributional properties of the underlying processes, namely the heavy tails. Estimation of the bivariate Hurst exponent and related separate Hurst exponents is thus a complex matter which needs to be approached and treated accordingly.

The results bring additional challenges to the time domain estimators construction, specifically its application under heavy tails. As the procedures are based on the absolute values of covariance scaling, the methods can confuse the scaling of covariances with scaling of tails of the separate processes. As shown by Barunik \& Kristoufek \cite{Barunik2010}, the time domain estimators are unbiased under heavy tails when estimating long-range dependence. When the setting is extended to the absolute values scaling, it very well corresponds to the $\alpha$ parameter of the $\alpha$-stable distributions. However, this could create a problem as the procedures might mistake the covariance absolute value scaling for the scaling of absolutes values of the separate processes, hence the upward bias reported in the simulation study. The remedy does not necessarily need to be too complicated as the heavy tails bias can be estimated using shuffling procedures or bootstrapping. This way, the time domain estimators could successfully compete with the frequency domain estimators even under heavy tails.

\section*{Acknowledgements}

The research leading to these results has received funding from the Czech Science Foundation under project No. 14-11402P.

\newpage


\bibliographystyle{unsrt}


\end{document}